\documentclass[%
 reprint,
 superscriptaddress,
 amsmath,amssymb,
 aps,
]{revtex4-2}

\usepackage{xcolor}
\usepackage{graphicx}
\usepackage{dcolumn}
\usepackage{lipsum}
\usepackage[caption=false]{subfig}

\usepackage[T1]{fontenc}
\usepackage[colorlinks=true]{hyperref}%

\newcommand{\norm}[1]{\left\lVert#1\right\rVert}
\def\Msun{{\mathrm{M}_{\odot}}}
\def\Mmax{{M_{\max}}}

\newcommand{\TEOB}{{\tt TEOBResumS}}
\newcommand{\red}[1]{\textcolor{red}{#1}}

\begin{document}

\preprint{APS/123-QED}

\title{Updated universal relations for tidal deformabilities\\of neutron stars from phenomenological equations of state}
 
\author{Daniel A. Godzieba}%
 \email{dag5611@psu.edu}
\affiliation{Department of Physics, The Pennsylvania
State University, University Park, Pennsylvania 16802}

\author{Rossella Gamba}
 \email{rossella.gamba@uni-jena.de}
 \affiliation{Theoretisch-Physikalisches Institut, Friedrich-Schiller-Universit{\"a}t Jena, 07743, Jena, Germany}

\author{David Radice}
 \email{dur566@psu.edu}
 \affiliation{Department of Physics, The Pennsylvania
State University, University Park, Pennsylvania 16802}
 \affiliation{Institute for Gravitation \& the Cosmos, The Pennsylvania State University, University Park, PA 16802}
 \affiliation{Department of Astronomy \& Astrophysics, The Pennsylvania State University, University Park, PA 16802}
 
 \author{Sebastiano Bernuzzi}
 \email{sebastiano.bernuzzi@uni-jena.de}
 \affiliation{Theoretisch-Physikalisches Institut, Friedrich-Schiller-Universit{\"a}t Jena, 07743, Jena, Germany}

\date{\today}

\begin{abstract}
Equation of state (EOS) insensitive relations, so-called universal relations, between the neutron star (NS) compactness, its multipolar tidal deformability coefficients, and between the tidal parameters for binary systems are essential to break degeneracies in gravitational wave data analysis. Here, we validate and recalibrate these universal relations using a large set of almost 2 million phenomenological EOSs that are consistent with current observations. In doing so, we extend universal relations to a larger region of the EOS parameter space, most notably to softer EOSs and larger compactnesses. We show that waveform models that neglect higher-than-leading-order tidal deformations of the NSs accumulate as much as $3.5$ radians of dephasing from $20\,{\rm Hz}$ to merger. We also perform a full Bayesian parameter estimation of the GW170817 data, and we compare the NS radius constraints produced using universal relations from the literature and the updated fits we propose here. We find that the new fits yield a NS radius that is smaller by about 500 meters. This difference is less than the statistical uncertainty on the radius at the signal-to-noise-ratio of GW170817, but it is significantly larger than the precision anticipated for next-generation detectors.
\end{abstract}

\maketitle
\section{\label{Introduction}Introduction}
There is still great uncertainty in the equation of state (EOS) that describes the incredibly dense nuclear matter of neutron stars (NSs) in the regime above nuclear saturation density ($\rho_\text{nuc} \simeq 2.7 \times 10^{14} \, {\rm g}/{\rm cm}^3$) due to the highly non-perturbative nature of nuclear matter in this regime. Consequently, there is great uncertainty in the properties of NSs predicted by theory that are highly dependent on the EOS, such as the maximum stable mass for a non-rotating NS ($\Mmax$), as well as the relation between the mass of an NS and its radius (a.k.a. the mass-radius curve). The collection of current NS mass measurements shows that the lower bound on the value of $\Mmax$ lies firmly within the range $1.9 - 2.0$ $\Msun$ \cite{Ozel:2016oaf,Lattimer:2012nd}, and the discovery of pulsar J0740+6620 ($M \simeq 2.14 \Msun$) strongly indicates that the lower bound could be constrained even higher \cite{Cromartie:2019kug}. Constraints on $\Mmax$ and measurements of NS radii have been combined to place constraints the EOS using both Bayesian/weighted \cite{Steiner:2010fz,Steiner:2012xt,Ozel:2015fia,Most:2018hfd,Capano:2019eae,Drischler:2020fvz,Most:2018hfd,Greif:2018njt} and unweighted \cite{Godzieba:2020tjn,Annala:2017llu,Annala:2019puf,De:2018uhw,Hebeler:2013nza} techniques. Upcoming precision NICER measurements of millisecond pulsar radii will likely constrain the EOS even further through these techniques \cite{Ozel:2016oaf,Godzieba:2020tjn}.

However, certain relations between bulk properties of NSs exhibit universality, meaning they are largely independent of the EOS. In the age of gravitational wave (GW) astronomy, some particularly important relations are those between the tidal deformability parameters (or, simply, tidal deformabilities) of NSs, which are related to the tidal Love numbers. During a binary neutron star (BNS) inspiral, the gravitational field of each star causes a deformation on the other star through tidal forces. These deformations, which are described by the tidal deformability parameters, alter the trajectory of each star, which becomes imprinted in the resultant GW signal. \citet{Yagi:2013sva} has demonstrated that a robust relation between several $l$-th order dimensionless electric tidal deformabilities, $\Lambda_l$, of non-rotating NSs exists across a variety of theoretical neutron star equations of state (EOS). \citet{Yagi:2015pkc,Yagi:2016qmr} have also shown that a similar relation exists for BNSs between the symmetric and antisymmetric combinations of each NS's electric quadrupolar tidal deformability, $\Lambda_2$. 

A universal relation reduces a group of parameters to a single parameter family; that is, the measurement of one parameter yields all others through the relation, breaking the degeneracy between them. In this analysis, we are concerned with a set of universal relations that are important for GW astronomy and LIGO/VIRGO observations (and have notably been used in the LIGO/VIRGO analysis of GW170817 \cite{TheLIGOScientific:2017qsa,Abbott:2018exr}).

First, there are universal relations between various multipole tidal deformabilities $\Lambda_l$, the so-called ``multipole Love relations.'' Tidal deformabilities enter into the waveform of GW signals of BNS inspirals at different post-Newtonian (PN) orders. (The $l$th order electric tidal deformability enters into the GW signal at $2l+1$ PN orders \cite{Yagi:2013sva}.) The dominant order is the quadrupole ($\Lambda_2$) term, followed by the much smaller octupole ($\Lambda_3$) and hexadecapole ($\Lambda_4$) terms. The $\Lambda_3$ and $\Lambda_4$ terms are difficult to measure accurately in the GW signal due to their small magnitudes and are often dropped to compute the leading order effect. However, the measurement of these quantities as well as the bias introduced by dropping them from calculations can be avoided entirely using universal relations. With the multipole relations, $\Lambda_3$ and $\Lambda_4$ can be computed directly using the more easily measureable $\Lambda_2$, leading to a manifold increase in measurement accuracy of $\Lambda_2$ \cite{Yagi:2013sva,Yagi:2016bkt}. These multipole relations, then, will be critical tools for the analysis of GW signals with upcoming third-generation GW detectors such as LIGO III and the Einstein Telescope \cite{Hinderer:2009ca,Yagi:2013sva}.

Next, there is a universal relation between $\Lambda_2$ and the compactness of a NS, $C \equiv M/R$, where $M$ and $R$ are the mass and radius of the NS respectively (we take $G=c=1$). This relation essentially falls out of the definition for $\Lambda_2$. The $l$-th order dimensionless electric tidal deformabiliy $\Lambda_l$ can be defined in terms of $C$ and the $l$-th order electric tidal Love number $k_l$ as \cite{Yagi:2013sva,Damour:2009vw} \begin{equation}
    \Lambda_l \equiv \frac{2}{(2l-1)!!} \frac{k_l}{C^{2l+1}}.
    \label{lambda definition}
\end{equation} We see also that the previous multipole relations follow from and have their physical origins in Eq. (\ref{lambda definition}). It can be shown that $k_2$ goes roughly as $C^{-1}$, independent of the EOS, over the range of $C$ values observed in Nature. Thus, overall, $\Lambda_2$ goes roughly as $C^{-6}$ for all EOSs \cite{Postnikov:2010yn,De:2018uhw}. There is a clear physical intuition for this relation. For a given NS mass $M$, the less compact the NS is (that is, the larger its radius $R$), the more easily it is deformed by a tidal potential, and thus the larger the value of $\Lambda_2$. This relation, then, allows one to convert constraints on $\Lambda_2$ from GW observations to constraints on the radius of the NS (or even to compute the radius directly from $\Lambda_2$) as has been done in the LIGO/VIRGO analysis of GW170817 \cite{Abbott:2018exr,De:2018uhw}.

Finally, there is a universal relation for BNSs between the symmetric and antisymmetric combinations of $\Lambda_2$ for each star, the so-called ``binary Love relation.'' Consider a NS binary with primary and secondary masses $m_{1}$ and $m_{2}$ ($m_{1} \geq m_{2}$) and respective quadrupolar tidal deformabilities $\Lambda_{2,1}$ and $\Lambda_{2,2}$. The symmetric and antisymmetric combinations of $\Lambda_{2,1}$ and $\Lambda_{2,2}$ are \begin{equation}
    \Lambda_{s} \equiv \frac{\Lambda_{2,1} + \Lambda_{2,2} }{2}, \quad \Lambda_{a} \equiv \frac{\Lambda_{2,2} - \Lambda_{2,1} }{2}. \label{sym and antisym definition}
\end{equation} The individual tidal deformabilites, $\Lambda_{2,1}$ and $\Lambda_{2,2}$, are degenerate in the GW phase information. What is actually measured in the GW signal is really a combination of $\Lambda_{2,1}$ and $\Lambda_{2,2}$ \cite{Yagi:2016bkt,Yagi:2016qmr}. Just as with the multipole relations, the relation between $\Lambda_{s}$ and $\Lambda_{a}$ (which also involves a third parameter, the binary mass ratio $q \equiv m_{2}/m_{1}$) reduces the analysis to the estimation and measurement of a single parameter, $\Lambda_{s}$, from which $\Lambda_{a}$ (and, thus, $\Lambda_{2,1}$ and $\Lambda_{2,2}$) can then be computed. Currently, this is the method by which Advanced LIGO is able to extract individual tidal deformability information from GW signals of BNSs \cite{Yagi:2016bkt,Yagi:2016qmr,Abbott:2018exr}. The approximate universality of the relation between $\Lambda_{a}$ and $\Lambda_{s}$ follows from the approximate no-hair relations for compact objects, arising from the approximate symmetry of isodensity self-similarity \cite{Yagi:2016bkt}.

In their original analysis, \citet{Yagi:2013sva} and \citet{Yagi:2015pkc} validated these universal relations against a set of a few very diverse theoretical EOS models, but not over the entire space of EOSs allowed by astronomical observation and theoretical calculations. The motivation for the work in this paper, then, is to validate these relations over a much broader extent of the space of all possible EOSs.

In this paper, we update and recalibrate the fits to these universal relations using a large set of randomly generated phenomenological EOSs that satisfy astronomical observation and theoretical calculations. The structure of this paper is as follows. In Sec. \ref{Methods} we will describe the parameterization of the four-piece polytrope EOS model and the algorithm by which the phenomenological EOSs are generated. In Sec.~\ref{Multipole and Compactness Relations}, we analyze the universal $\Lambda_3$-$\Lambda_2$, $\Lambda_4$-$\Lambda_2$, and $C$-$\Lambda_2$ relations from the collection of phenomenological EOSs. We present the fitting parameters of these relations and compare them to previous fits. In Sec.~\ref{Binary Relations}, we analyze the $\Lambda_a$-$\Lambda_s$ relation and compare the fitting parameters to previous fits, including the fit currently used by the LIGO/VIRGO collaboration. In Sec.~\ref{GWapp}, we discuss the applications of the updated fits to GW modelling and parameter estimation. A concluding summary is given in Sec.~\ref{Conclusion}.

\section{\label{Methods}Methods}

\subsection{\label{Parameterization}EOS Parameterization}
In order to explore the space of all possible EOSs that satisfy known observational constraints and theoretical calculations (known as the EOS band), we employ a Markov chain Monte Carlo (MCMC) algorithm that generates random piecewise polytropic EOSs. We use a variation of the piecewise polytropic interpolation scheme developed by \citet{Read:2008iy}. The scheme models the EOS as a continuous piecewise function of four polyropes: \begin{equation}
    p(\rho) = \begin{cases} K_0 \rho^{\Gamma_0} \quad \rho \leq \rho_0 \\ K_1 \rho^{\Gamma_1} \quad \rho_0 < \rho \leq \rho_1 \\ K_2 \rho^{\Gamma_2} \quad \rho_1 < \rho \leq \rho_2 \\ K_3 \rho^{\Gamma_3} \quad \rho > \rho_2. \end{cases}
    \label{EOS}
\end{equation} A four-piece model allows for a great diversity of EOSs (ex. hard/soft EOSs, EOSs with/without phase transitions, etc.) and ensures that the most extreme regions of the EOS band will be reached by the MCMC algorithm. The specific choice of a piecewise polytropic ansatz for the EOS, as opposed to known alternative schemes, does not significantly bias the resultant shape of the computed EOS band \cite{Annala:2019puf}.

The first polytrope piece of the model corresponds to the presumed known EOS of the outer and inner crust up to around nuclear density, where $K_0 = 3.59389\times10^{13}$~[cgs] and $\Gamma_0 = 1.35692$ \cite{Douchin:2001sv}. Here, nuclear density is taken to be $\rho_\text{nuc} = 2.7 \times 10^{14}$ g/cm$^3$. This piece is fixed for all randomly generated EOSs. The specific choice of the low-density crust EOS does not significantly determine the bulk physical properties of the NS \cite{Read:2008iy,Rhoades:1974fn,Most:2018hfd}

The $K_i$ are determined by continuity; thus, the last three polytrope pieces of the EOS are specified by six parameters: three transition densities ($\rho_0$, $\rho_1$, and $\rho_2$) and three adiabatic indices ($\Gamma_1$, $\Gamma_2$, and $\Gamma_3$). The authors of \cite{Read:2008iy} reduce this to four parameters by fixing the values of $\rho_1$ and $\rho_2$, as an EOS with a smaller number of parameters can be reasonably constrained by a only a few astronomical observations. However, this imposes a prior assumption on the form of the EOS and narrows the parameter space to a much smaller region of the EOS band. Thus, to probe the entire EOS band (including extreme EOSs not ruled out by observation) without imposing assumptions about the true form of the EOS, we allow $\rho_1$ and $\rho_2$ to be free parameters as well. Therefore, each EOS is specified by the full set of six parameters: $\rho_0 \in [0.15\rho_\text{nuc},1.2\rho_\text{nuc}]$, $\rho_1 \in [1.5\rho_\text{nuc},8\rho_\text{nuc}]$, $\rho_2 \in (\rho_1,8.5\rho_\text{nuc}]$, $\Gamma_1 \in [1.4,5]$, $\Gamma_2 \in [0,8]$, $\Gamma_3 \in [0.5,8]$.

Though continuity is imposed on each EOS, the speed of sound within the NS ($c_{s}$) as a function of density for each EOS is not necessarily continuous at the transition densities. \citet{OBoyle:2020qvf} have recently developed a modified version of the piecewise polytropic scheme by \citet{Read:2008iy} used in this analysis which imposes continuity on $c_{s}$. We do not use this modified scheme, as it was published after our analysis. However, \citet{Kanakis-Pegios:2020jnf} have shown that the effects of discontinuities in $c_{s}$ on the bulk properties of a NS are negligible.

\subsection{\label{MCMC Algorithm}MCMC Algorithm}
To probe the EOS band in a way that is both thorough and computationally efficient, we use a MCMC algorithm in the form of a random walk through the parameter space. The constraints of the EOS band define a path-connected region of the six-dimensional parameter space. A path between any two points in this region, then, corresponds to a smooth deformation of the EOS at one point to the EOS at the second point. Therefore, a series of small, random deformations of the parameterized EOS model would correspond to a random walk through the parameter space. This is the basic idea behind the algorithm.

Taking the logarithm of Eq. (\ref{EOS}) converts the EOS to a piecewise linear function. A deformation can then be performed very easily by shifting the positions of just four points: the three transition points $\textbf{r}_0 = \{\log(\rho_0),\log(p_0)\}$, $\textbf{r}_1 = \{\log(\rho_1),\log(p_1)\}$, and $\textbf{r}_2 = \{\log(\rho_2),\log(p_2)\}$; and an endpoint $\textbf{r}_3 = \{15.5,\log(p_3)\}$. The density value of the endpoint is kept fixed at $\rho = 10^{15.5}$ g/cm$^3$, but this choice is arbitrary, as the only purpose of the endpoint is to define the slope of the last linear piece, $\Gamma_3$. Both $\textbf{r}_0$ and $\textbf{r}_3$ are restricted to move one-dimensionally ($\textbf{r}_0$ can only shift along the crust EOS line and $\textbf{r}_3$ can only shift vertically up or down), while $\textbf{r}_1$ and $\textbf{r}_2$ have the full two degrees of freedom. Thus, the whole EOS still has six degrees of freedom.

\begin{figure}
	\centering
	\includegraphics[trim=10 5 20 20, clip, width=\linewidth]{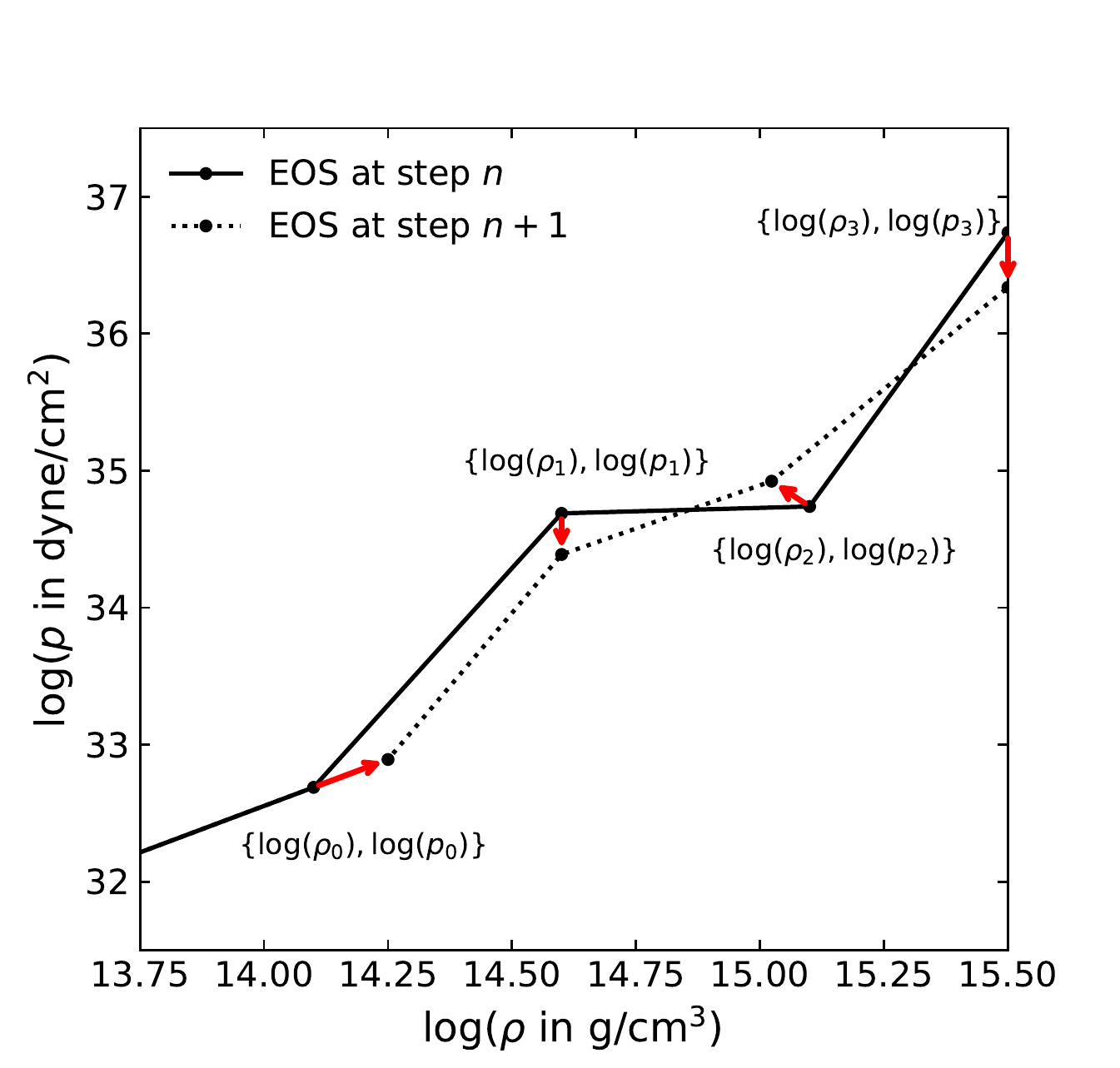}
	\caption{A single step of the MCMC algorithm. Each of the four points used to define the current EOS (solid line) is independently shifted by a displacement vector of random magnitude (shown in red), creating a trial EOS (dashed line). Trial EOSs are repeatedly generated from the current EOS until one is found that satisfies observational constraints. When this occurs, the current EOS is updated to the trial EOS and the process repeats.}
	\label{Deformation}
\end{figure}

A single step of the MCMC algorithm proceeds as follows and is illustrated in Fig. \ref{Deformation}.  The current EOS is defined by the four points $\textbf{r}_i$ and corresponds to the current position of the algorithm in parameter space. For each $\textbf{r}_i$, a displacement vector $\Delta \textbf{r}_i$ is independently generated from a uniform distribution with a random direction (respecting the point's degrees of freedom) and a random magnitude (up to a maximum size $\norm{\Delta \textbf{r}_i} \leq 0.05$). A trial EOS is then defined by the four new points, $\textbf{r}_i' = \textbf{r}_i + \Delta\textbf{r}_i$. The transition densities and adiabatic indicies of this trial EOS are then checked to see if they are within the bounds given in Sec. \ref{Parameterization}.

The physical properties of the trial EOS are then found by computing a sequence of solutions to the Tolman-Oppenheimer-Volkoff (TOV) equation. We utilize the publicly available \texttt{TOVL} code described in \citet{Bernuzzi:2008fu} and \citet{Damour:2009vw} to solve the TOV equation. The trial EOS is accepted if it satisfies three weak physical constraints that define the EOS band:\begin{enumerate}
    \setlength\itemsep{0em}
    \item causality of the maximum mass NS (i.e. the sound speed is subluminal, $c_s < c$);
    \item $\Mmax > 1.97 \, \Msun$;
    \item $\Lambda_{2} < 800$ for the $1.4 \, \Msun$ NS.
\end{enumerate} The upper limit on  $\Lambda_{2}$ is the 90\%-credible upper bound derived in the LIGO/VIRGO analysis of GW170817 \cite{TheLIGOScientific:2017qsa}. If the trial EOS is accepted, its parameters and physical properties are then recorded, and the current EOS is updated to the new EOS, $\textbf{r}_i \rightarrow \textbf{r}_i'$.

The initial EOS is randomly selected from a set of EOSs that satisfy the constraints computed via a standard Monte Carlo analysis of the parameter space. The algorithm is then allowed to proceed until a specified number of steps have been completed.

\begin{figure*}[t]
    \centering
    \subfloat[]{{\includegraphics[trim=10 0 18 20, clip, width=0.45\linewidth]{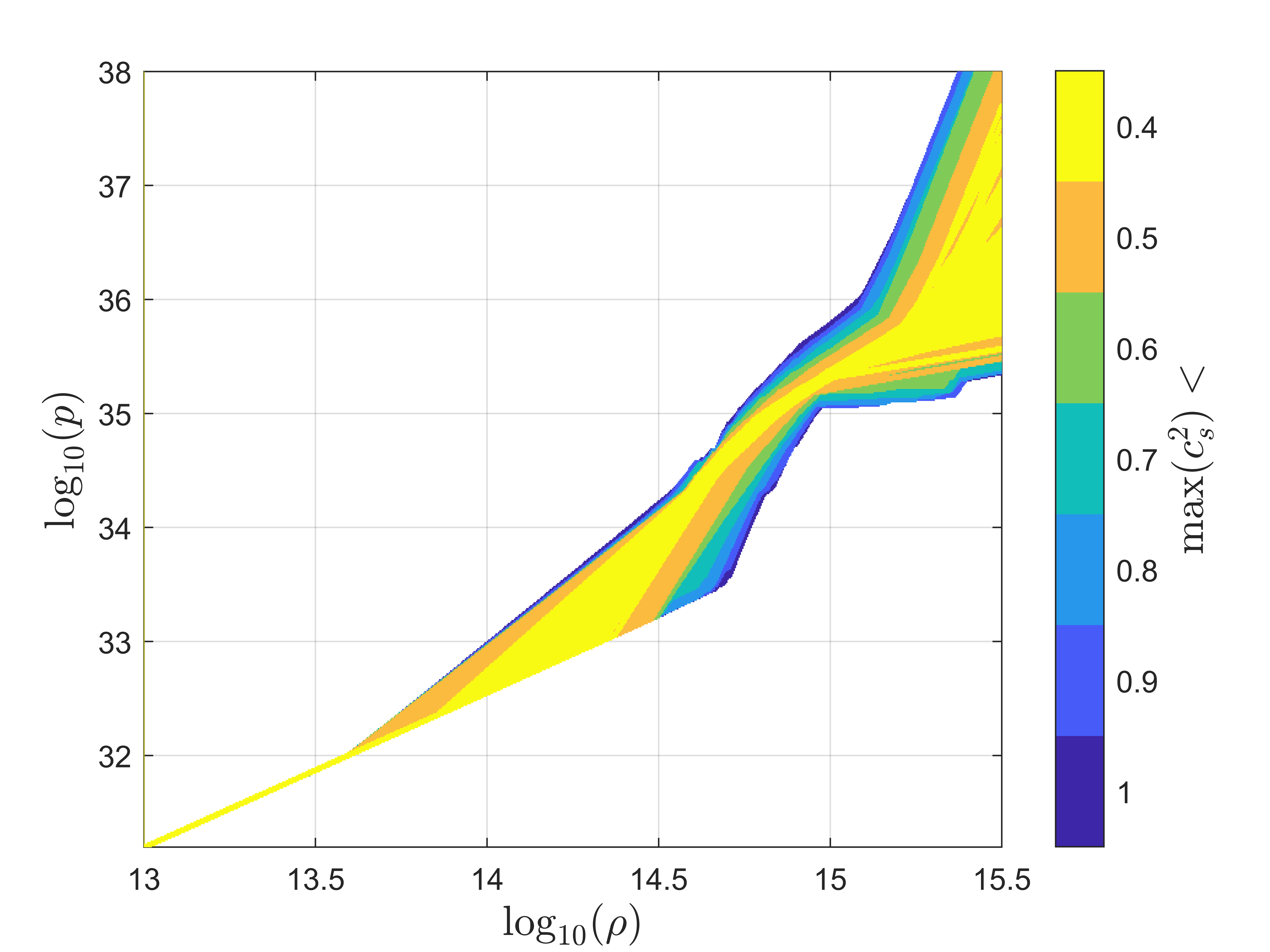}}}%
    \qquad
    \subfloat[]{{\includegraphics[trim=10 0 18 20, clip, width=0.45\linewidth]{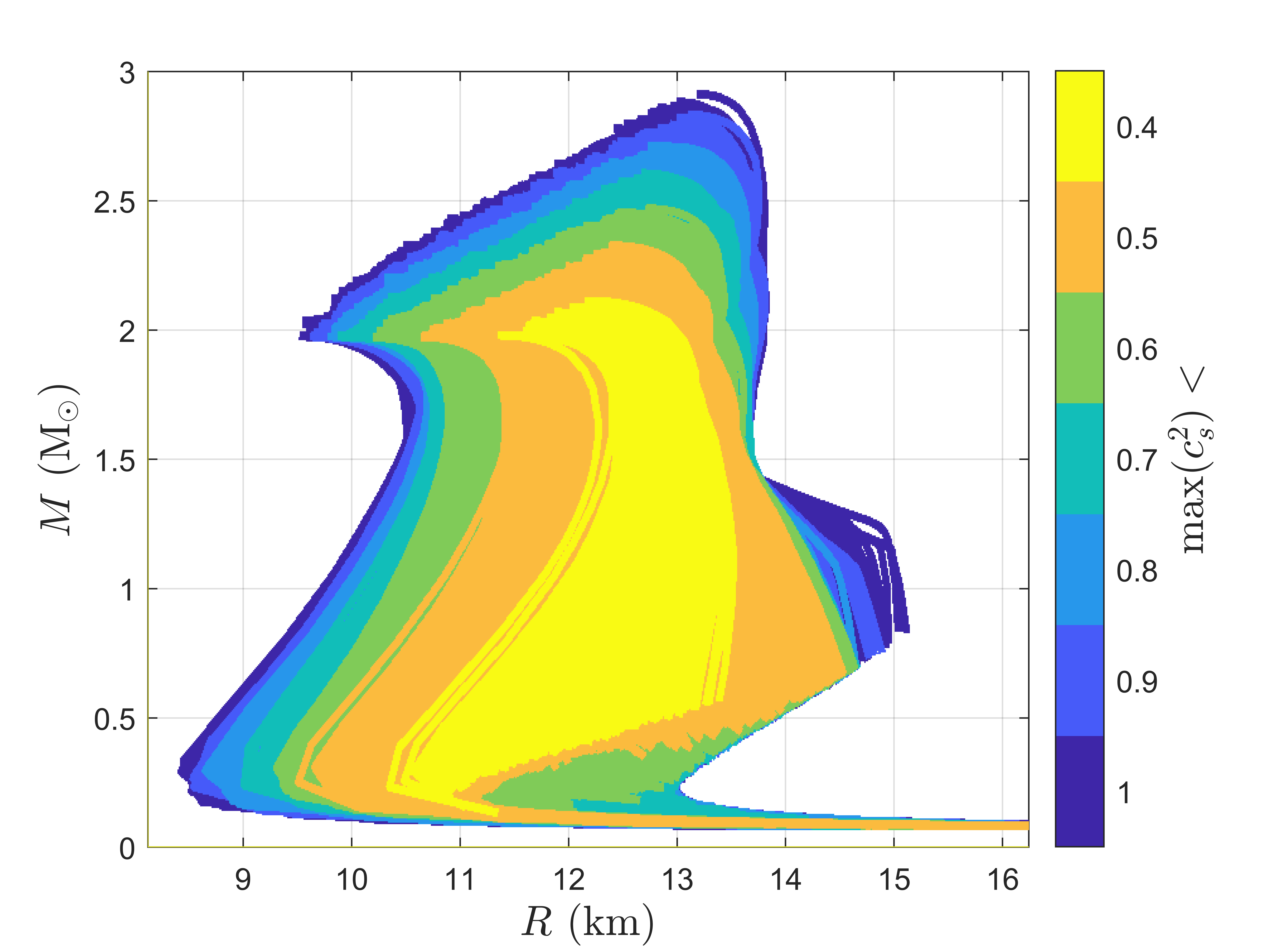}}}%
    \caption{The collection of 1,966,225 phenomenological EOSs (a) and associated mass-radius curves (b)  computed using the MCMC algorithm. Color here is used to indicate the maximum sound speed $c_s$ reached within the maximum mass NS of each EOS (EOSs with smaller $\max(c_{s}^{2})$ are drawn on top of ones with larger $\max(c_{s}^{2})$). The collection reveals the approximate shape of the EOS band. The most extreme EOSs that reach the edges of the band are those where $c_s=1$.}%
    \label{EOS band}%
\end{figure*}

\section{\label{Multipole and Compactness Relations}Multipole and Compactness Relations}

\begin{figure*}[t]
    \centering
    \subfloat[]{{\includegraphics[trim=0 5 20 20, clip, width=0.4\linewidth]{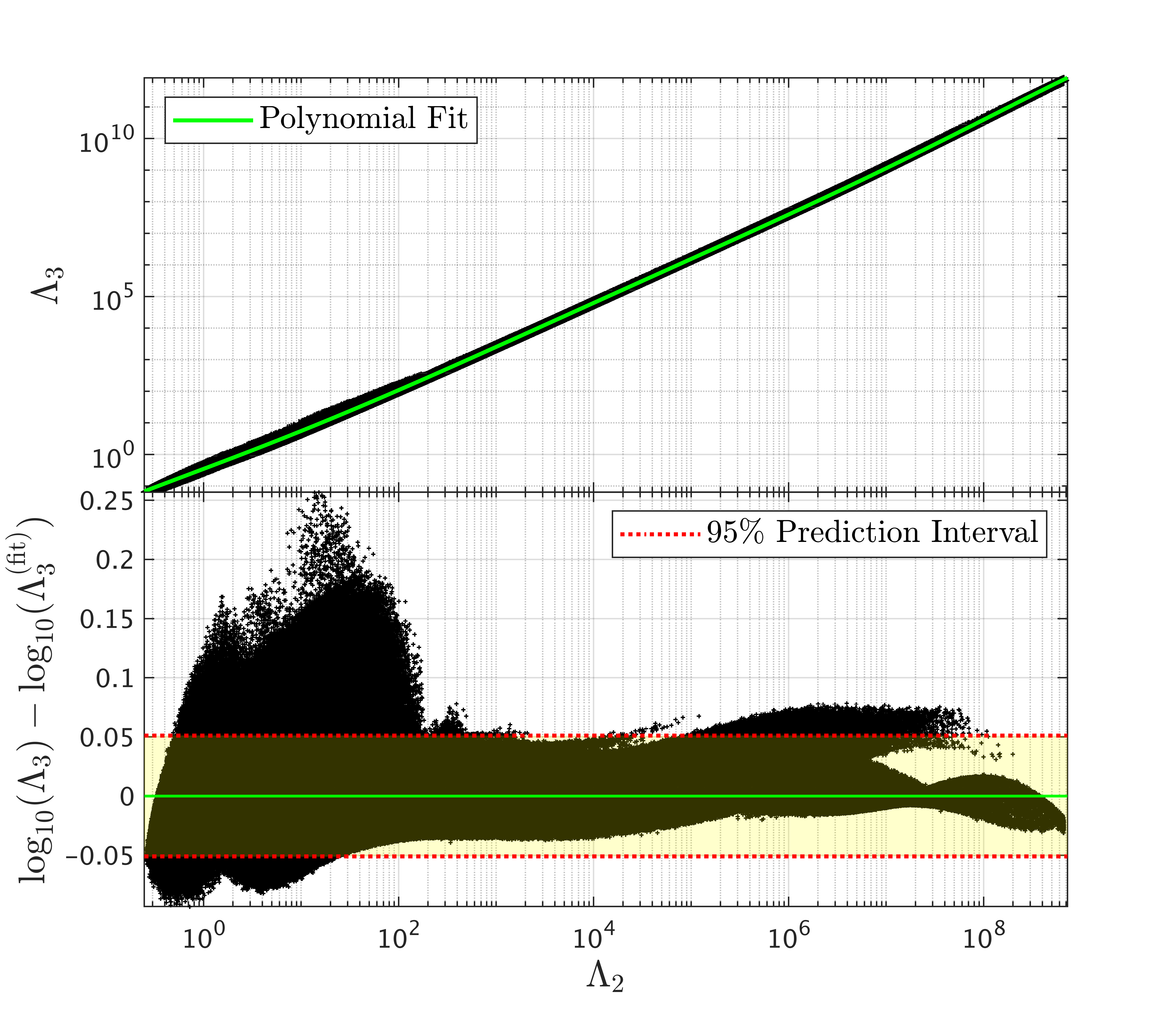}}}%
    \qquad
    \subfloat[]{{\includegraphics[trim=0 5 20 20, clip, width=0.4\linewidth]{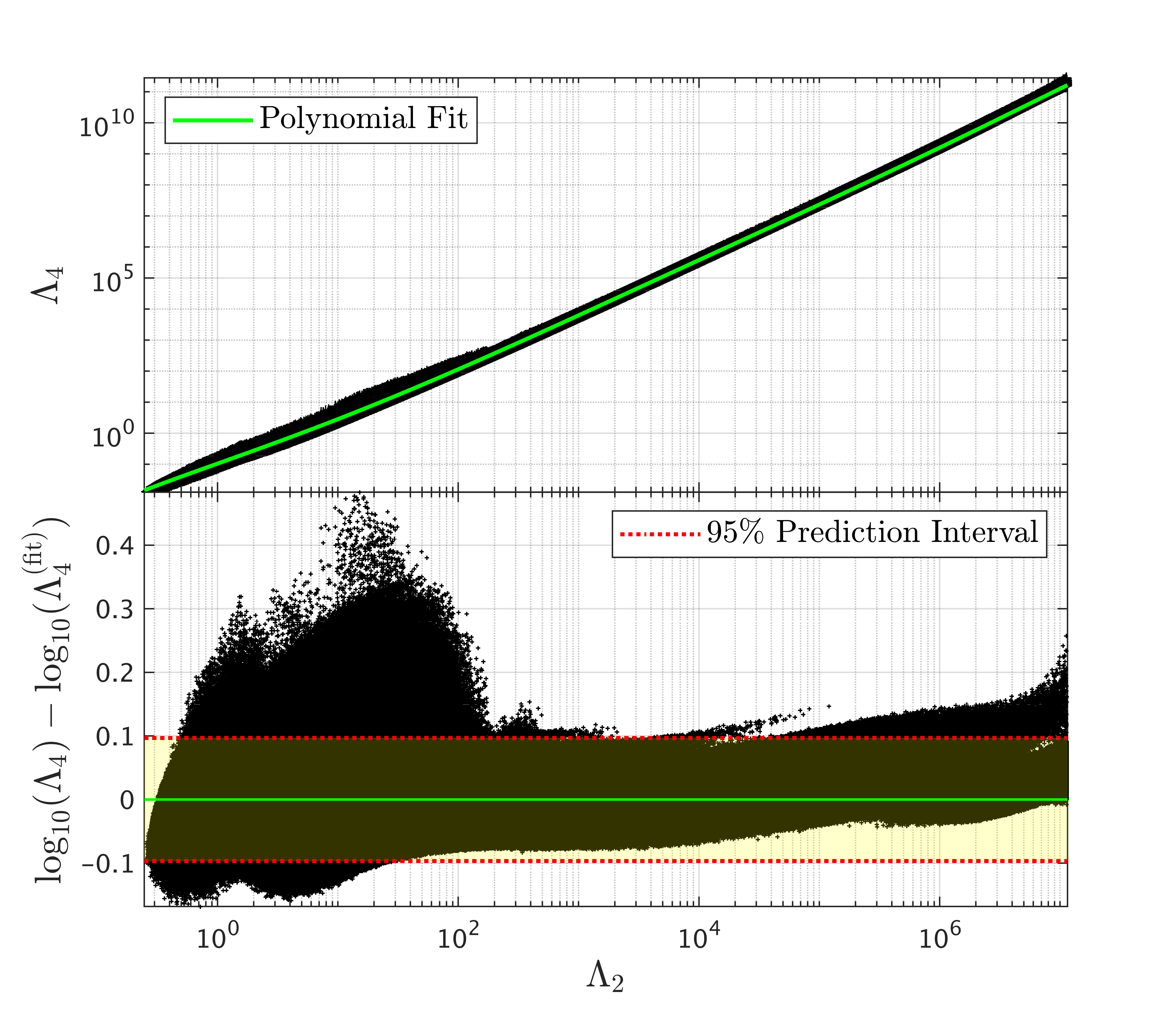}}}%
    \caption{Universal (a) $\Lambda_3$-$\Lambda_2$ and (b) $\Lambda_4$-$\Lambda_2$ relations for NSs from the collection of phenomenological EOSs. Sixteen NSs with central densities in the range $5.0\times10^{-4} \ {\rm g}/{\rm cm}^3 \leq \rho_c \leq 5.0\times10^{-3} \ {\rm g}/{\rm cm}^3$ were computed for each EOS. The relations are fitted with the polynomial expression in Eq. (\ref{multipole fit}), with the fitting parameters for each relation given in Table \ref{fitting parameters 1}. The log residuals of the both fits are shown with 95\% prediction intervals.}%
    \label{multipole fit plots}%
\end{figure*}

\begin{figure}
	\centering
	\includegraphics[trim=0 0 0 0, clip, width=\linewidth]{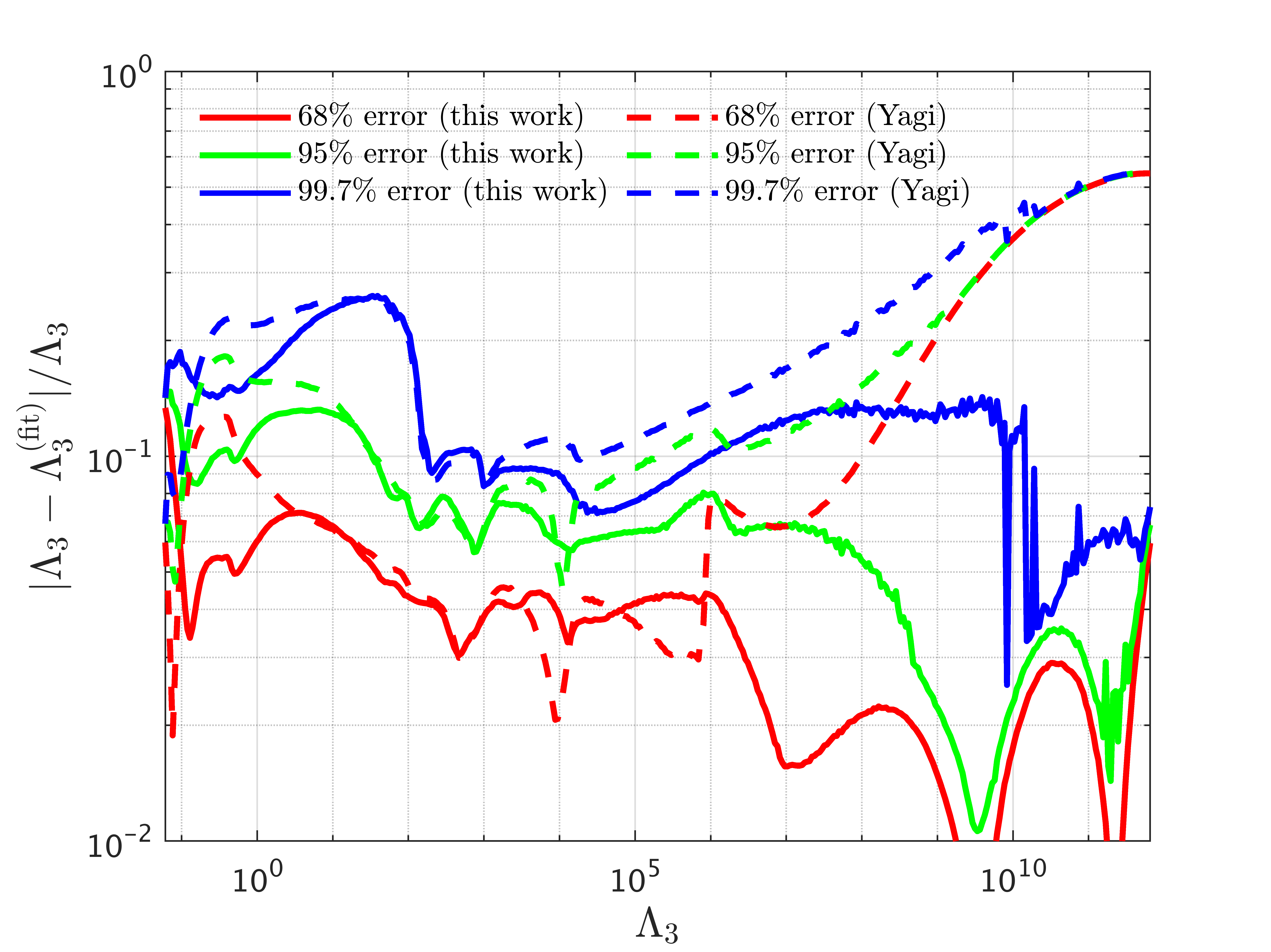}
	\caption{68\%, 95\%, and 99.7\% relative error of the $\Lambda_3$-$\Lambda_2$ fit as a function of $\Lambda_3$. The solid lines represent the error of the new fit in this work, and the dashed lines represent the error of the original fit by \citet{Yagi:2013sva}.}
	\label{relative error 3}
\end{figure}

\begin{figure}
	\centering
	\includegraphics[trim=0 0 0 0, clip, width=\linewidth]{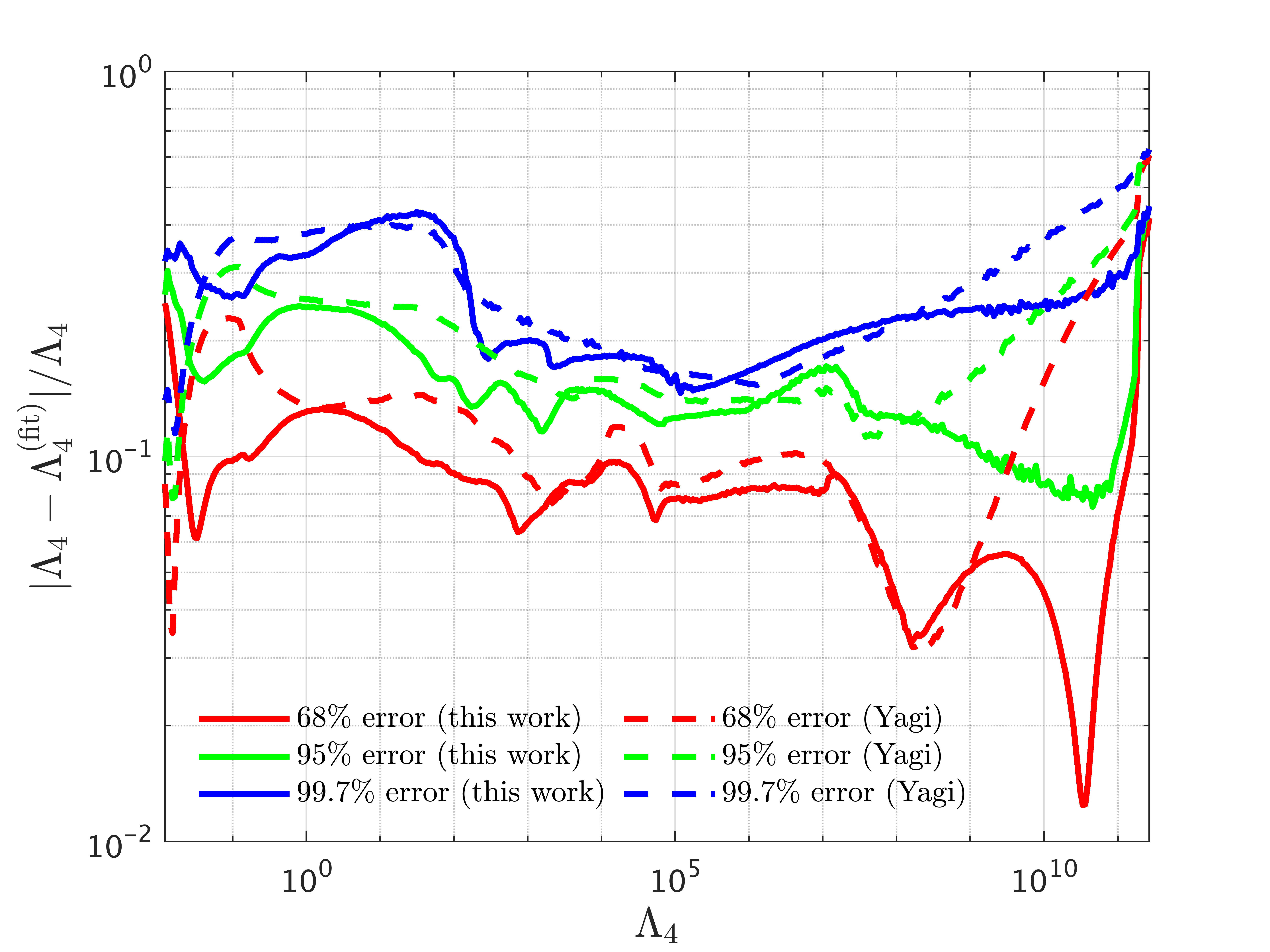}
	\caption{68\%, 95\%, and 99.7\% relative error of the $\Lambda_4$-$\Lambda_2$ fit as a function of $\Lambda_4$. The solid lines represent the error of the fit in this work, and the dashed lines represent the error of the original fit by \citet{Yagi:2013sva}.}
	\label{relative error 4}
\end{figure}

\begin{figure}
	\centering
	\includegraphics[trim=0 0 0 0, clip, width=\linewidth]{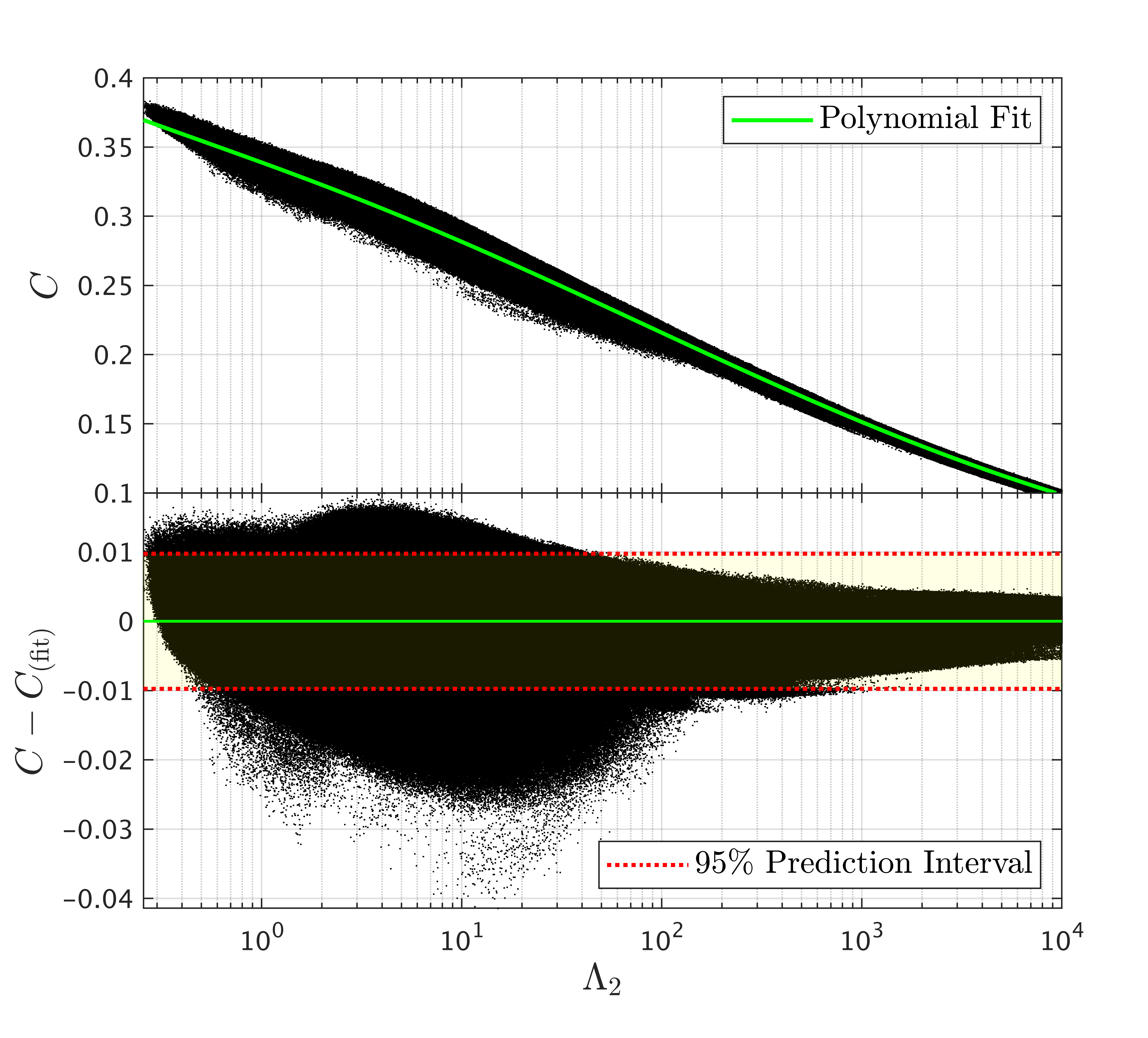}
	\caption{Universal $C$-$\Lambda_2$ for NSs from the collection of phenomenological EOSs. The relation is fitted with the expression in Eq. (\ref{C-lambda2 fit}), with the fitting parameters given in Table \ref{fitting parameters 2}. The residual of the fit is shown with the 95\% prediction interval. The distribution of $\Lambda_2$ values roughly goes as $C^{-6}$.}
	\label{C-lambda2 plot}
\end{figure}

\begin{figure}
	\centering
	\includegraphics[trim=5 0 20 17, clip, width=\linewidth]{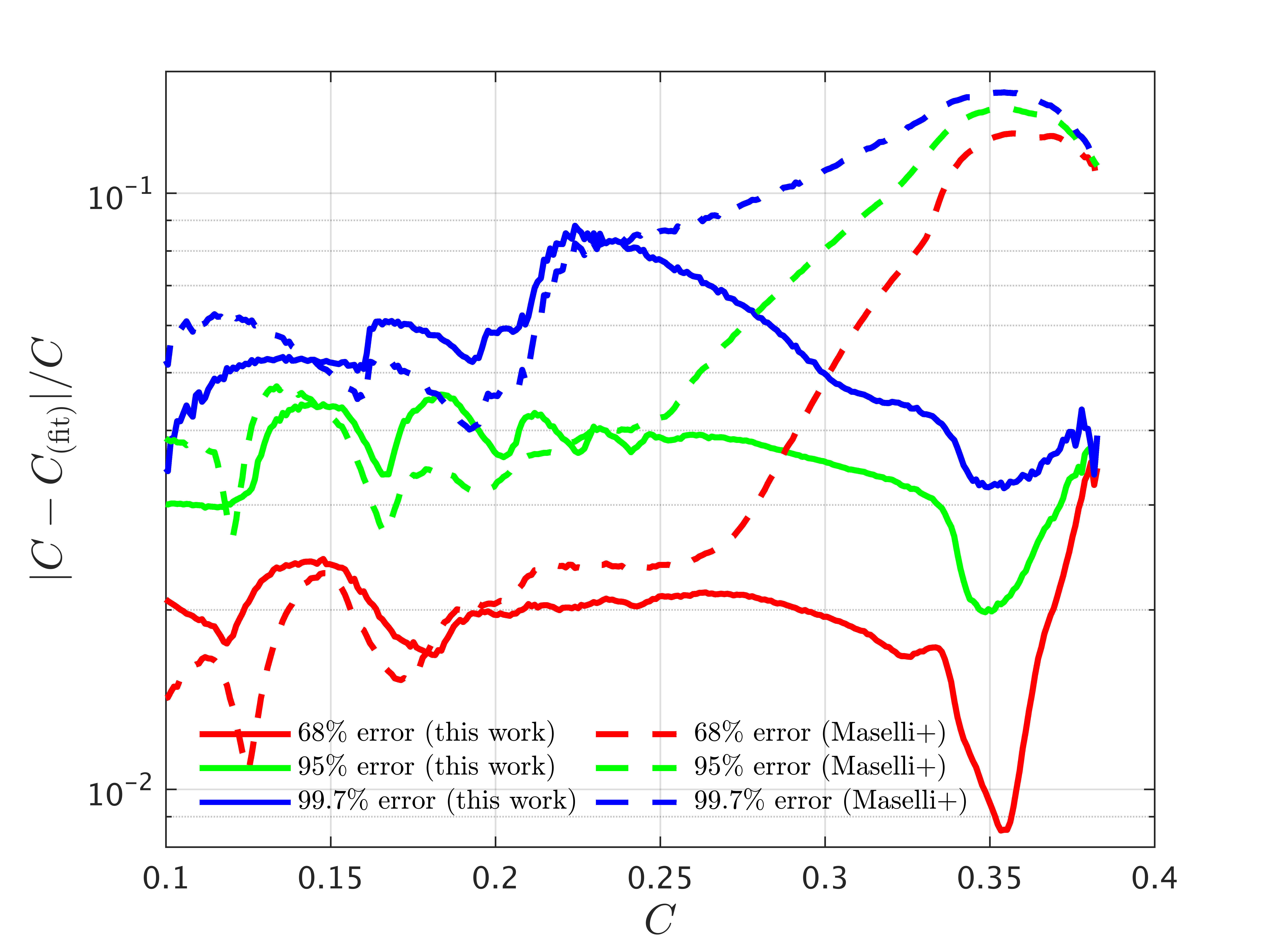}
	\caption{68\%, 95\%, and 99.7\% relative error of the $C$-$\Lambda_2$ fit as a function of $C$. The solid lines represent the error of the fit in this work, and the dashed lines represent the error of the original fit by \citet{Maselli:2013mva}. Our fit has a drastically smaller error at larger values of $C$ than the original fit has.}
	\label{C_lambda2 relative error}
\end{figure}

Using the MCMC algorithm, we generate a set of 1,966,225 phenomenological EOSs. The complete set of EOSs is plotted in Fig. \ref{EOS band} along with the associated mass-radius curve for each EOS. The plot reveals the approximate shape of the full EOS band defined by the three week constraints in Sec. \ref{MCMC Algorithm}. The edges of the band are populated by the most extreme EOSs where $c_{s}=1$. The upper limit for the value of $\Mmax$ correlates strongly with $\Lambda_2$ for the 1.4 $\Msun$ NS \cite{Godzieba:2020tjn}; consequently, the constraint that $\Lambda_2 < 800$ for the 1.4 $\Msun$ NS effectively functions as an upper constraint on the value of $\Mmax$. The largest value of $\Mmax$ in our data set is 2.9096 $\Msun$, below the theoretical upper limit from GR of 3.2 $\Msun$ \cite{Rhoades:1974fn}, which can be seen in the mass-radius band in Fig. \ref{EOS band}. Similarly, and quite understandably, the upper limit for the radius of the 1.4 $\Msun$ NS, $R_{1.4}$, also correlates strongly with $\Lambda_2$ for the 1.4 $\Msun$ NS \cite{Godzieba:2020tjn}. The $\Lambda_2 < 800$ constraint then also functions as an upper limit for $R_{1.4}$. This explains why a kink can be seen on the right side of the mass-radius band at $M \sim 1.4 \ \Msun$ in Fig. \ref{EOS band}. The largest value of $R_{1.4}$ in our data set is 13.9 km.
 
The universal $\Lambda_3$-$\Lambda_2$, $\Lambda_4$-$\Lambda_2$, and $C$-$\Lambda_2$ relations are the nearly-EOS-independent relations that reduce $\Lambda_3$, $\Lambda_4$, and $C$ to functions of the single parameter $\Lambda_2$ for any given NS. To analyze these relations across our set of phenomenological EOSs, we solve the TOV equation for each EOS at sixteen evenly spaced central density values in the range $5\times10^{-4} \ {\rm g}/{\rm cm}^3 \leq \rho_{c} \leq 5\times10^{-3} \ {\rm g}/{\rm cm}^3$ and then extract $\Lambda_2$, $\Lambda_3$, $\Lambda_4$, and $C$ from the NS solution at each density. (If the central density of the maximum mass NS for a given EOS falls below any of the density values, then the TOV equation is not solved for those density values.) A total of 30,464,895 NS solutions were computed.

In Fig. \ref{multipole fit plots}, we plot (a) $\Lambda_3$ vs. $\Lambda_2$  and (b) $\Lambda_4$ vs. $\Lambda_2$ for our set of NS solutions. The $\Lambda_3$-$\Lambda_2$ and $\Lambda_4$-$\Lambda_2$ relations are fitted with the polynomial-like expression \begin{equation}
    \ln{\Lambda_{3,4}} = \sum_{k=0}^{6} a_{k} \left( \ln{\Lambda_{2}} \right)^{k}.
    \label{multipole fit}
\end{equation} This is an updated version of the fitting function used by \citet{Yagi:2013sva}. The original fitting function is quartic in $\ln{\Lambda_{2}}$. However, with our larger data set, a quartic fit is insufficient to remove trends from the residual of each relation. Thus, we employ a fit that is 6th order in $\ln{\Lambda_{2}}$.  The quality of the fits can be appreciated in Fig. \ref{multipole fit plots}, where the residuals of $\log_{10}{\Lambda_{3}}$ and $\log_{10}{\Lambda_{4}}$ are given with 95\% prediction intervals. In Figs. \ref{relative error 3} and \ref{relative error 4}, we compare the the 68\%, 95\%, and 99.7\% relative errors of the fits in this work to those the original fits by \citet{Yagi:2013sva}. At each percentage error level, our fits demonstrate a general improvement in accuracy over the original fits.

Over the range of tidal parameters most relevant to LIGO observations, $\Lambda_2,\Lambda_3,\Lambda_4 \in [1,10^4]$, our $\Lambda_3$-$\Lambda_2$ fit holds to a maximum error of $\sim$30\%, with 95\% of the errors below $\sim$13\%. The original fit to this relation holds to a maximum error of $\sim$30\%, with 95\% of the error below $\sim$14\%. Our $\Lambda_4$-$\Lambda_2$ fit holds to a maximum error of $\sim$45\%, with 95\% of the errors below $\sim$25\%. The original fit to this relation holds to a maximum error of $\sim$40\%, with 95\% of the error below $\sim$26\%.

The fitting parameters $\vec{a} = \{a_k\}$ of the new fits and the original quartic fits by \citet{Yagi:2013sva} are given in Table \ref{fitting parameters 1}. The leading-order terms of the fits are the constant ($k=0$) and linear ($k=1$) terms. (This is evident from the near-linear shape of the distribution of points in each log-log plot in Fig. \ref{multipole fit plots}.) Therefore, a direct comparison can be made between the leading-order terms of tbe original fits and our new fits. The coefficients $a_0$ and $a_1$ for our fits are in good agreement with those of the original fits. This demonstrates the validity of the original $\Lambda_3$-$\Lambda_2$ and $\Lambda_4$-$\Lambda_2$ fits across the EOS band. 
 
Just like the multipole relations, the universal relation between $C$ and $\Lambda_2$ reduces $C$ to a function of $\Lambda_2$. This relation essentially falls out of the definition of $\Lambda_2$ in Eq. (\ref{lambda definition}). In Fig. \ref{C-lambda2 plot}, we plot $C$ vs. $\Lambda_2$ for our set of NS solutions. The $C$-$\Lambda_2$ relation is fitted with a similar polynomial-like expression \begin{equation}
    C = \sum_{k=0}^{6} a_{k} \left( \ln{\Lambda_{2,3,4}} \right)^{k}
    \label{C-lambda2 fit}
\end{equation} This fitting function is an updated version of the function used by \citet{Maselli:2013mva} and presented in \cite{Yagi:2016qmr}. The original fitting function is quadratic in $\ln{\Lambda_2}$. Once again, with our larger data set, we are required to go to 6th order in $\ln{\Lambda_2}$ to remove all trends from the residual of our fit. The fitting parameters $\vec{a} = \{a_k\}$ are also given in Table \ref{fitting parameters 1}. 

In Fig. \ref{C_lambda2 relative error}, we compare the the 68\%, 95\%, and 99.7\% relative errors of the fit in this work to those of the original fit by \citet{Maselli:2013mva}. Our fit demonstrates a significant improvement in accuracy over the original fit at larger values of $C$.

As one can see from Fig. \ref{C-lambda2 fit}, the range of $\Lambda_2$ values most relevant to LIGO observations, $\Lambda_2 \in [1,10^4]$, translates into a corresponding range $C \in [0.1,0.35]$. Over this range, our fit holds to a maximum error of $\sim$10\%, with 95\% of the errors below $\sim$5\%. The original fit to this relation holds to a maximum error of $\sim$15\%, with 95\% of the error below $\sim$14\%. This is the largest improvement in accuracy out of all the fits in this work. We are able then to confirm the concerns raised by \citet{Kastaun:2019bxo} that existing fits to the $C$-$\Lambda_2$ are unreliable at large $C$.

The fitting parameters of the quadratic fit by \citet{Maselli:2013mva} are given in Table \ref{fitting parameters 2}. The leading-order terms are once again the constant and linear terms, as can be seen in the near-linear shape of the distribution of points in Fig. \ref{C-lambda2 plot}. The coefficients $a_0$ and $a_1$ are in generally good agreement between the original and updated fits; however, there is a relatively large difference between the values of $a_1$ as compared to what is seen with the multipole fits. This is a consequence of the improvements to the fit made by using both a higher-order polynomial function and a larger data set over a greater extent of the EOS band.

In addition to the our fit to the $C$-$\Lambda_2$ relation, we also present entirely new fits to the $C$-$\Lambda_3$ and $C$-$\Lambda_4$ relations, the fitting parameters of which are also given in Table \ref{fitting parameters 2}. While these two relations can be derived from the multipole and $C$-$\Lambda_2$ fits, these explicit fits have a smaller error, comparable to that of the $C$-$\Lambda_2$ fit. 

\begin{table*}[t]
  \centering
  \begin{tabular}{ccccccccc}
    \hline \hline
    Fit & Relation & $a_{0}$ & $a_{1}$ & $a_{2}$ & $a_{3}$ & $a_{4}$ & $a_{5}$ & $a_{6}$ \\
    \hline
    \citet{Yagi:2013sva} & $\Lambda_3$-$\Lambda_2$  & $-1.15$ & $1.18$ & $2.51\times10^{-2}$ & $-1.31\times10^{-3}$ & $2.52\times10^{-5}$ & - & - \\
         & $\Lambda_4$-$\Lambda_2$  & $-2.45$ & $1.43$ & $3.95\times10^{-2}$ & $-1.81\times10^{-3}$ & $2.80\times10^{-5}$ & - & - \\
    This work & $\Lambda_3$-$\Lambda_2$  & $-1.052$ & $1.165$ & $6.590\times10^{-3}$ & $4.990\times10^{-3}$ & $-7.176\times10^{-4}$ & $3.741\times10^{-5}$ & $-6.694\times10^{-8}$ \\
         & $\Lambda_4$-$\Lambda_2$  & $-2.260$ & $1.384$ & $2.845\times10^{-4}$ & $1.287\times10^{-2}$ & $-1.856\times10^{-3}$ & $1.041\times10^{-4}$ & $-2.080\times10^{-6}$ \\
    \hline \hline
  \end{tabular}
  \caption{Fitting parameters $\vec{a} = \{a_k\}$ for the $\Lambda_3$-$\Lambda_2$ and $\Lambda_4$-$\Lambda_2$ relations given in Eq. (\ref{C-lambda2 fit}) from the original fits in \cite{Yagi:2013sva} and from the fits in this work to the phenomelogical EOSs. }
  \label{fitting parameters 1}
\end{table*}

\begin{table*}[t]
  \centering
  \begin{tabular}{ccccccccc}
    \hline \hline
    Fit & Relation & $a_{0}$ & $a_{1}$ & $a_{2}$ & $a_{3}$ & $a_{4}$ & $a_{5}$ & $a_{6}$ \\
    \hline
    \citet{Maselli:2013mva} & $C$-$\Lambda_2$  & $0.371$ & $-3.91\times10^{-2}$ & $1.056\times10^{-3}$ & - & - & - & - \\
    This work & $C$-$\Lambda_2$  & $0.3388$ & $-2.30\times10^{-2}$ & $-4.651\times10^{-4}$ & $-2.636\times10^{-4}$ & $5.424\times10^{-5}$ & $-3.188\times10^{-6}$ & $6.181\times10^{-8}$ \\
         & $C$-$\Lambda_3$  & $0.3180$ & $-2.066\times10^{-2}$ & $-4.627\times10^{-4}$ & $3.421\times10^{-5}$ & $3.694\times10^{-6}$ & $-2.673\times10^{-7}$ & $4.626\times10^{-9}$ \\
         & $C$-$\Lambda_4$  & $0.3001$ & $-1.764\times10^{-2}$ & $-1.518\times10^{-4}$ & $1.442\times10^{-5}$ & $2.239\times10^{-6}$ & $-1.492\times10^{-7}$ & $2.523\times10^{-9}$ \\
    \hline \hline
  \end{tabular}
  \caption{Fitting parameters $\vec{a} = \{a_k\}$ for the $C$-$\Lambda_2$, $C$-$\Lambda_3$, and $C$-$\Lambda_4$ relations given in Eq. (\ref{C-lambda2 fit}) from the original fit in \cite{Maselli:2013mva} and from the fits in this work to the phenomelogical EOSs. }
  \label{fitting parameters 2}
\end{table*}

\section{\label{Binary Relations}Binary Relation}

\begin{figure*}[t]
    \centering
    \subfloat{{\includegraphics[trim=20 20 20 20, width=0.4\linewidth]{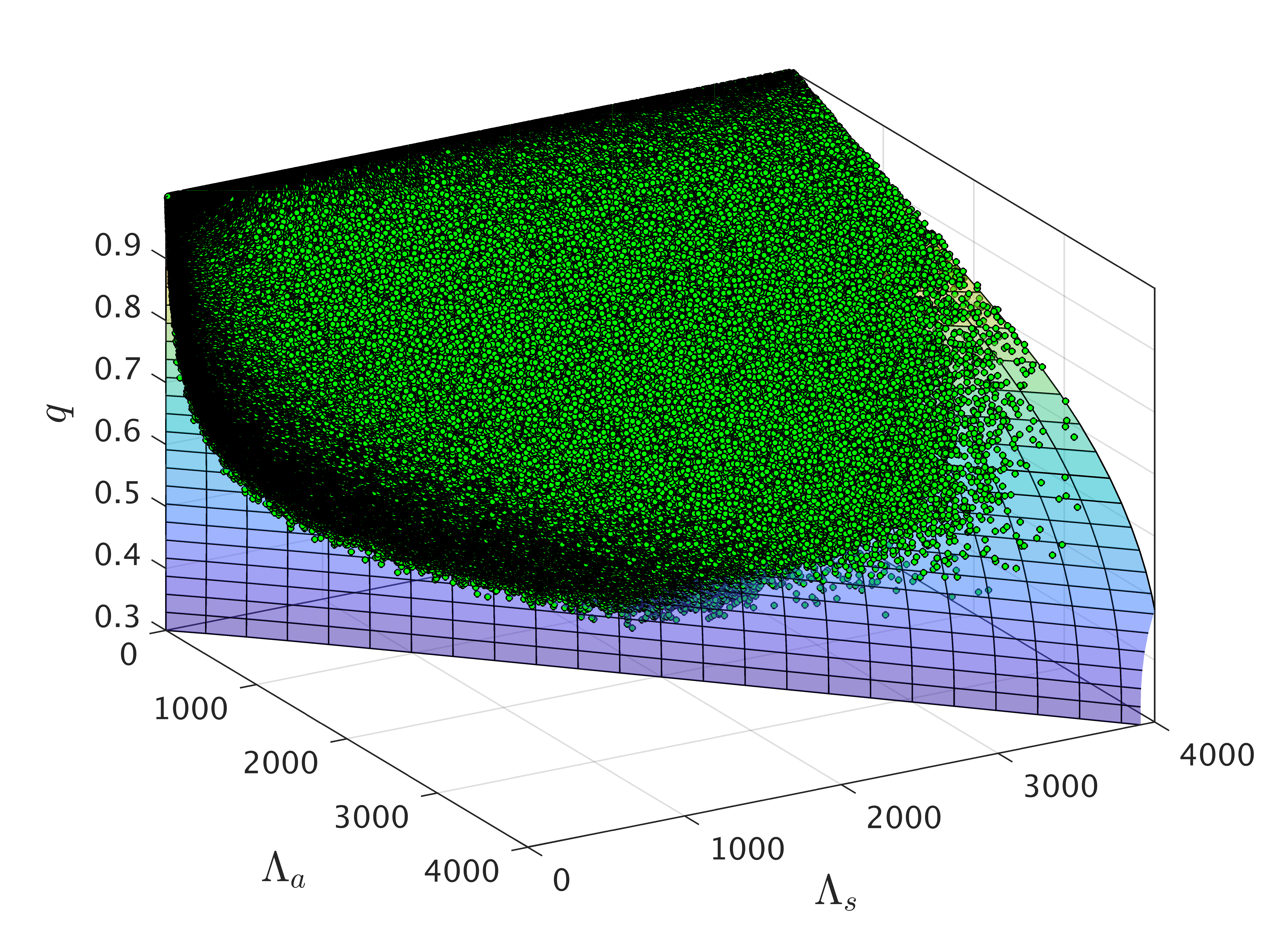} }}%
    \qquad
    \subfloat{{\includegraphics[trim=20 20 20 20, width=0.4\linewidth]{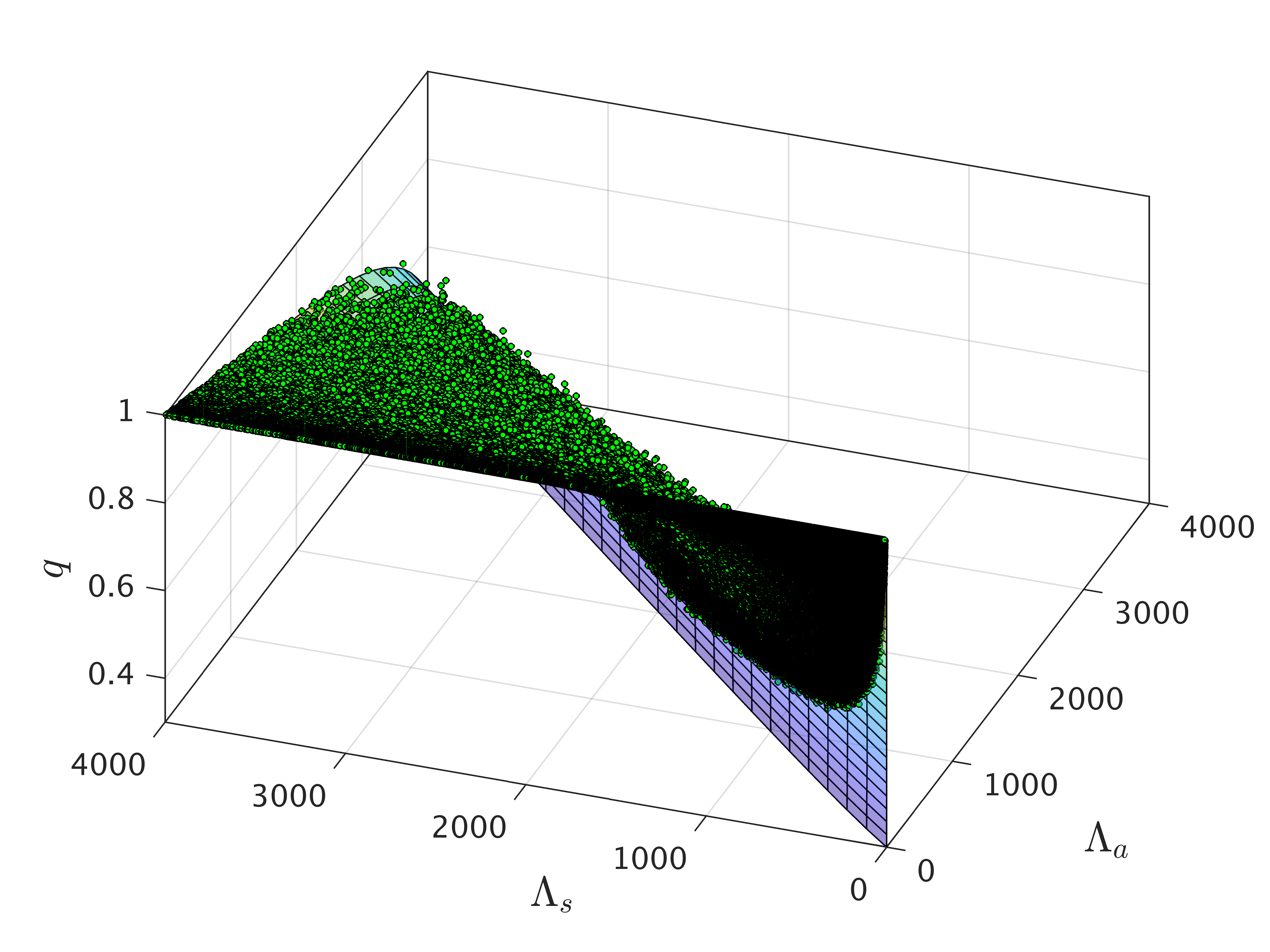} }}%
    \caption{Approximate universal $\Lambda_{a}$-$\Lambda_{s}$ relation for NSs from the collection of phenomenological EOSs, shown from two different angles. Twenty random binaries with $1 \, \Msun \leq m_2 \leq m_1 \leq \Mmax$ were computed for each EOS. The relation was fitted with the expression in Eq. (\ref{binary fit}), with the fitting parameters given in Table \ref{fitting parameters 3} 2}%
    \label{binary fit plot}%
\end{figure*}

\begin{figure}
	\centering
	\includegraphics[trim=0 0 0 0, clip, width=\linewidth]{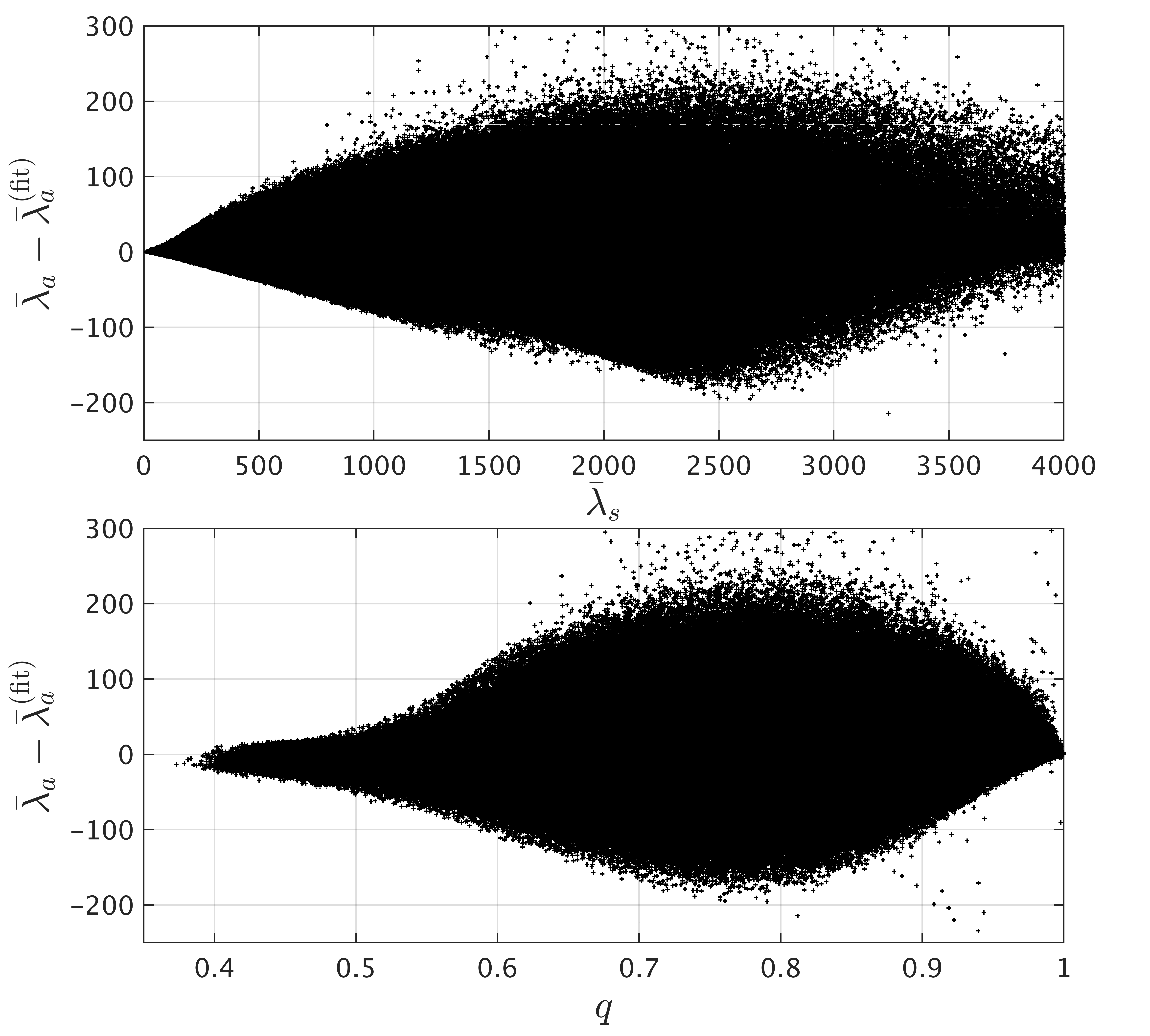}
	\caption{Residual of the $\Lambda_a$-$\Lambda_s$ relation fit for $\Lambda_s < 4000$ seen from two perspectives. The fit becomes increasingly better as $\Lambda_s \rightarrow 0$ and as $q \rightarrow 1$.}
	\label{binary fit residual}
\end{figure}

\begin{figure}
	\centering
	\includegraphics[trim=5 0 20 17, clip, width=\linewidth]{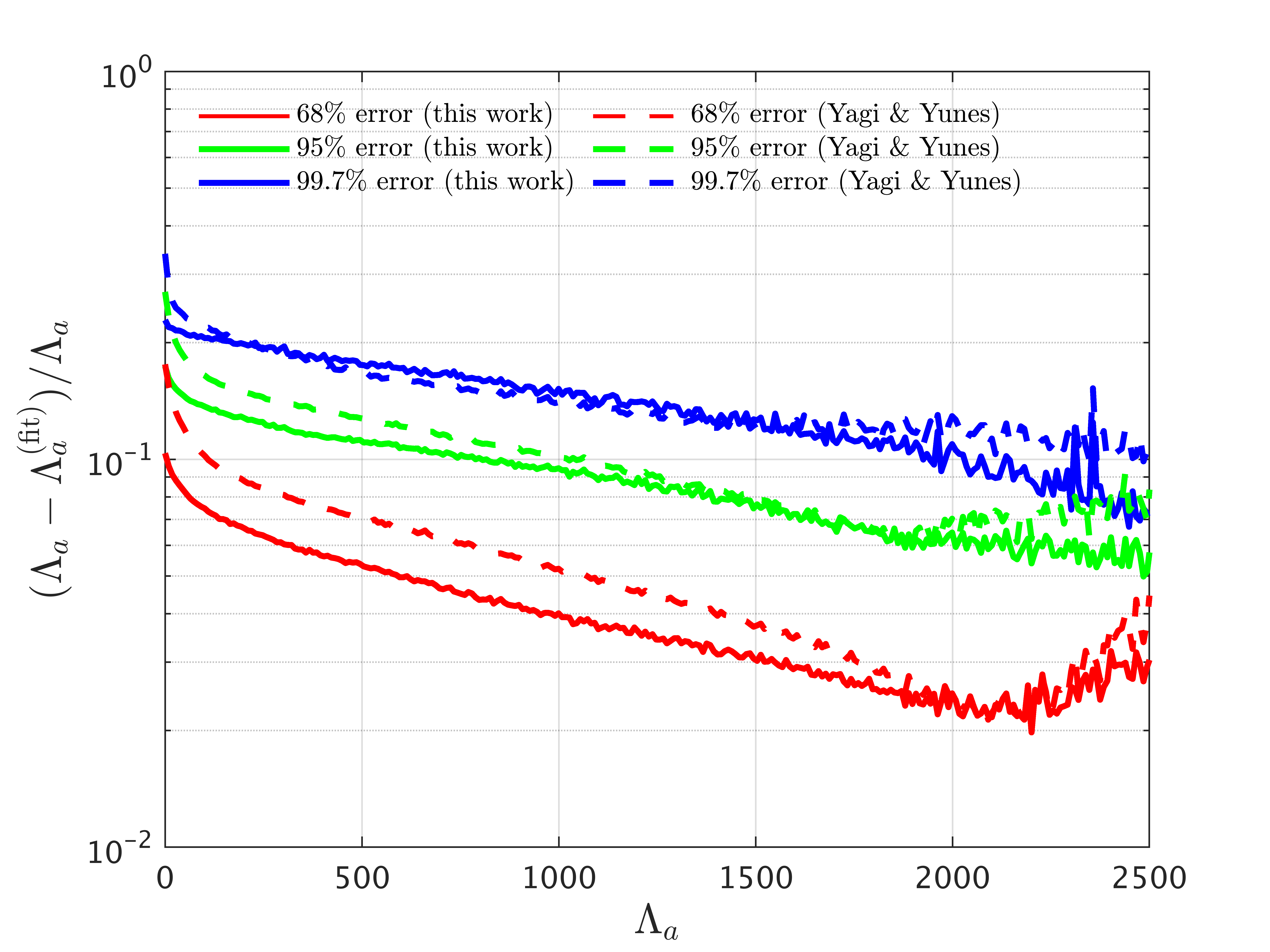}
	\caption{68\%, 95\%, and 99.7\% relative error of the $\Lambda_a$-$\Lambda_s$ fit as a function of $\Lambda_a$. The solid lines represent the error of the fit in this work, and the dashed lines represent the error of the original fit by \citet{Yagi:2015pkc,Yagi:2016qmr}.}
	\label{binary relative error}
\end{figure}

\begin{figure}[t]
    \includegraphics[trim=10 0 20 17, clip, width=\linewidth]{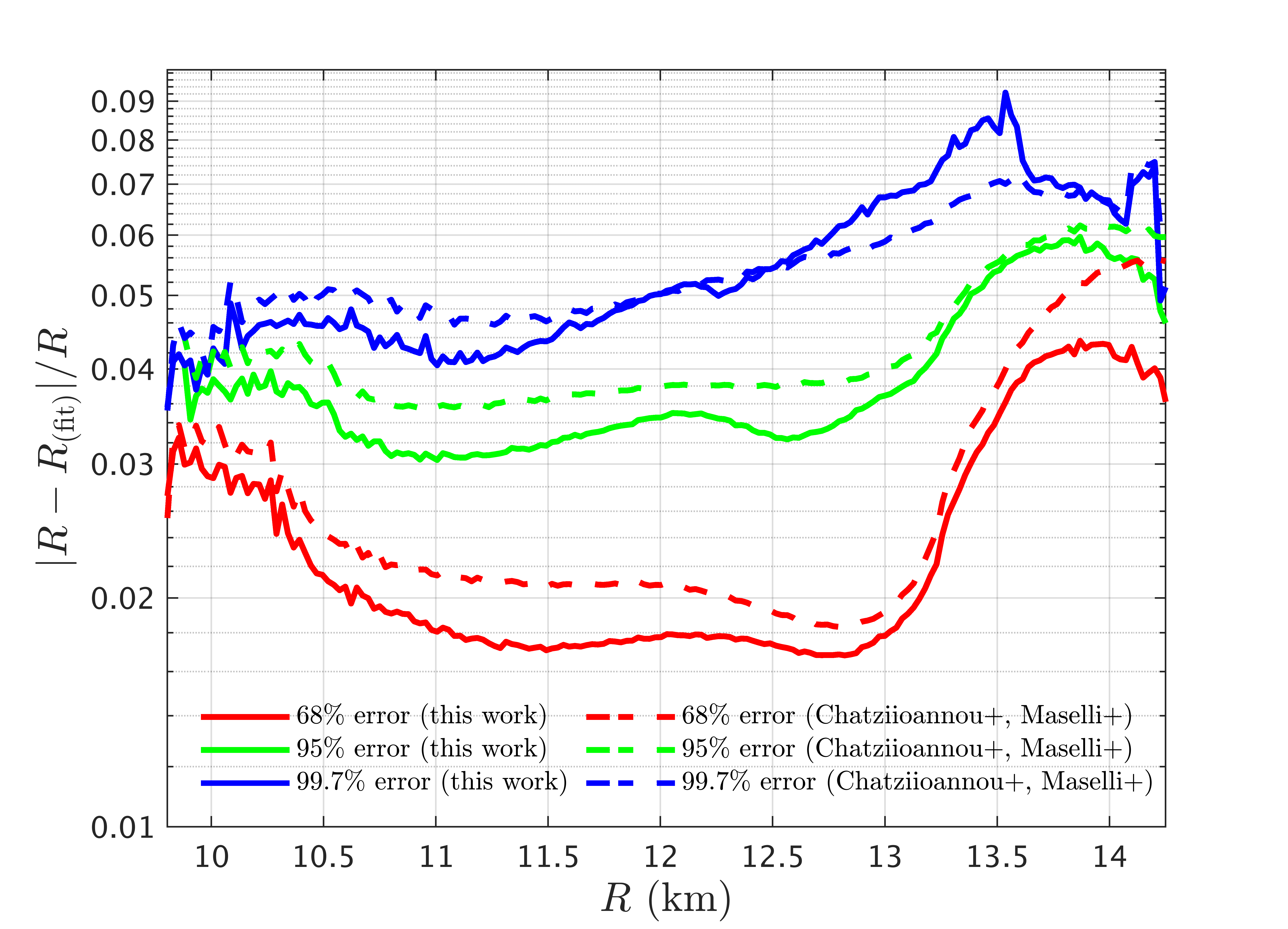}%
    \caption{68\%, 95\%, and 99.7\% relative error of NS radii as a recovered using the fits to the $\Lambda_a$-$\Lambda_s$ and $C$-$\Lambda_2$ relations as a function of the radius. The solid lines represent the error of as a result of the fits in this work, and the dashed lines represent the error as a result of the fits by \citet{Chatziioannou:2018vzf} and \citet{Maselli:2013mva}.}%
    \label{radius error comparison}%
\end{figure}

\begin{table*}[t]
  \centering
  \begin{tabular}{cccccccccccccc}
    \hline \hline
    Fit & $b_{11}$ & $b_{12}$ & $b_{21}$ & $b_{22}$ & $b_{31}$ & $b_{32}$ & $c_{11}$ & $c_{12}$ & $c_{21}$ & $c_{22}$ & $c_{31}$ & $c_{32}$ \\
    \hline
    \citet{Yagi:2015pkc,Yagi:2016qmr}\footnote{The parameter $a$ from these papers has been factored out of these parameter values.}  & -29.60 & 11.22 & 138.41 & -43.06 & -207.95 & 180.26 & -27.13 & 7.915 & 105.18 & 7.494 & -97.48 & -17.48 \\
    \citet{Chatziioannou:2018vzf} & -27.74 & 8.42 & 122.69 & -19.76 & -175.50 & 133.71 & -25.56 & 5.585 & 92.03 & 26.86 & -70.25 & -56.31 \\
    This work & -18.32 & 3.875 & 28.06 & 11.08 & 43.56 & 17.3 & -18.37 & 1.338 & 15.99 & 55.07 & 98.56 & -135.1 \\
    \hline \hline
  \end{tabular}
  \caption{Fitting parameters $\vec{b} = \{b_{ij},c_{ij}\}$ for the $\Lambda_a$-$\Lambda_s$ relation given in Eq. (\ref{binary fit}) from the original fit in \cite{Yagi:2015pkc,Yagi:2016qmr}, from the updated fit in \cite{Chatziioannou:2018vzf} used by LIGO/VIRGO, and from the fit in this work to the phenomenological EOSs. For all three fits, the average effective polytropic index is taken to be $\bar{n}=0.743$.}
  \label{fitting parameters 3}
\end{table*}

The universal $\Lambda_{a}$-$\Lambda_{s}$ relation is slightly more complicated  than the multipole and compactness relations, as it reduces $\Lambda_{a}$ to a function of two parameters: $\Lambda_{s}$ and the binary mass ratio $q$ for any given NS. We analyze the $\Lambda_{a}$-$\Lambda_{s}$ relation across the set of phenomenological EOSs by computing sequences of random BNSs using a random sample of a quarter of the total number of EOSs. We use the convention that the primary and secondary masses are given by $m_1$ and $m_2$ respectively, so $m_1 \geq m_2$. The binary mass ratio is then defined as $q \equiv m_2/m_1$. For each EOS, twenty random BNSs were generated. For each BNS, two masses were selected uniformly from the range $[1 \ \Msun, \Mmax]$, where $\Mmax$ is the maximum stable mass for the given EOS. The larger mass and smaller mass were then set as $m_1$ and $m_2$ respectively. The TOV equation was then solved for both stars in the binary. The quadrupolar tidal deformability of each star was extracted from the solution. Here we define $\Lambda_{2,1}$ and $\Lambda_{2,2}$ as the tidal deformabilities of the primary and secondary respectively. The symmetric and antisymmetric combinations of $\Lambda_{2,1}$ and $\Lambda_{2,2}$ were then computed using the definitions in Eq. (\ref{sym and antisym definition}). A total of 5,454,778 BNS solutions were computed.

In Fig. \ref{binary fit plot}, we plot $\Lambda_{a}$ vs. $\Lambda_{s}$ vs. $q$ for our set of BNS solutions. We employed the same fitting function used in the original analysis by \citet{Yagi:2015pkc,Yagi:2016qmr}, which is a Pad\'e approximant multiplied by a controlling factor:\begin{equation}
    \Lambda_a = F_{\bar{n}}(q) \, \Lambda_s \, \frac{1 + \sum_{i=1}^{3} \sum_{j=1}^{2} b_{ij} q^j \Lambda_s^{i/5}}{1 + \sum_{i=1}^{3} \sum_{j=1}^{2} c_{ij} q^j \Lambda_s^{i/5}}.
    \label{binary fit}  
\end{equation} The controlling factor $F_n(q)$ is derived from the Newtonian limit of the $\Lambda_{a}$-$\Lambda_s$ relation where the EOS is treated as a single Newtonian polytrope with polytropic index $n$: \begin{equation}
    F_{n}(q) \equiv \frac{1 - q^{10/(3-n)}}{1 + q^{10/(3-n)}}.
    \label{F(q)}
\end{equation} If one computes this limit with an EOS that is \textit{not} a single polytrope, $n$ instead represents the \textit{effective} polyropic index of the EOS. When considering \textit{multiple} EOSs in the context of the universal relation, $n$ is replaced with $\bar{n}$, which is the \textit{average} effective polytropic index across the set of EOSs. \citet{Yagi:2015pkc} originally analyzed the $\Lambda_{a}$-$\Lambda_{s}$ relation using a set of theoretical EOSs with an average effective polytropic index of $\bar{n} = 0.743$. \citet{Chatziioannou:2018vzf} reproduced the fit using three theoretical EOSs with diverse physical properties while still using $\bar{n} = 0.743$. The reproduced fit is the version that was  used in the LIGO/VIRGO analysis of GW170817 \cite{Abbott:2018exr}. In this analysis, we also take $\bar{n} = 0.743$ to allow for a direct comparison between our fit and the two previous fits. The fitting parameters $\vec{b} = \{b_{ij},c_{ij}\}$ for all three fits are given in Table \ref{fitting parameters 3}. We use the parameters of the fit by \citet{Chatziioannou:2018vzf} as the initial parameter values for our fit.

The residual of the fit is shown from two perspectives in Fig. \ref{binary fit residual}. In Fig. \ref{binary relative error}, we compare the the 68\%, 95\%, and 99.7\% relative errors of the this fit to those the original fit by \citet{Yagi:2015pkc,Yagi:2016qmr}. (We verify that the relative error of the fit by \citet{Chatziioannou:2018vzf} is identical to that of the original fit.) Our fit holds to a maximum error of $\sim$40\%, with 95\% of the error below $\sim$29\%. The original fit holds to a maximum error of ${\sim}43\%$, with 95\% of the error below ${\sim}31\%$.

We demonstrate the improvement our updated fits provide over the original fits to GW analysis by comparing the accuracy with which the radii of NS can be recovered from GW signals. The reduced tidal parameter \begin{equation}
    \tilde \Lambda = \frac{16}{13} \left(  \frac{(m_1 + 12m_2)m_1^4 \Lambda_{2,1}}{(m_1+m_2)^5} + \frac{(m_2 + 12m_1)m_2^4 \Lambda_{2,2}}{(m_1+m_2)^5} \right)
    \label{lambda tilde}
\end{equation} can be recovered the GW waveform \cite{Favata:2013rwa,Flanagan:2007ix,Radice:2020ddv} and is used by LIGO to compute the radii of both NSs in a binary merger. First, we compute the value of $\tilde \Lambda$ for each BNS solution using Eq. (\ref{lambda tilde}). Using the definitions in Eq. (\ref{sym and antisym definition}) and the $\Lambda_{a}$-$\Lambda_{s}$ relation, we can re-express Eq. (\ref{lambda tilde}) in terms of $\Lambda_{s}$ and $q$. Taking $m_1$ and $m_2$ to be known, we then solve Eq. (\ref{lambda tilde}) numerically for $\Lambda_{s}$ and recover a measurement of $\Lambda_{s}$ for each BNS solution. The $\Lambda_{a}$-$\Lambda_{s}$ relation then yields a measurement for $\Lambda_{a}$, allowing us to recover $\Lambda_{2,1}$ and $\Lambda_{2,2}$. Finally, we use the $C$-$\Lambda_{a}$ to compute $C$ for each NS, which when combined with $m_1$ and $m_2$ gives us the radius of each NS. In Fig. \ref{radius error comparison},  we compare the the 68\%, 95\%, and 99.7\% relative errors of the radii recovered using the fits in this work to those recovered using the fits by \citet{Chatziioannou:2018vzf} and \citet{Maselli:2013mva}. At the 68\% and 95\% error levels especially, our recovery show a definite improvement in accuracy over the original recovery. This is primarily a result of the drastic improvement in accuracy afforded by the updated $C$-$\Lambda_{s}$ fit. Our recovery has a maximum error of $\sim$10\%, with 95\% of the of the error below $\sim$6\%. The original recovery has a maximum error of $\sim$8\%, with 95\% of the error below $\sim$6.5\%

\section{\label{GWapp}Application to waveform modeling and PE}

\begin{figure*}
	\centering
	\subfloat[]{\includegraphics[trim=0 0 50 50, clip, width=0.49\linewidth]{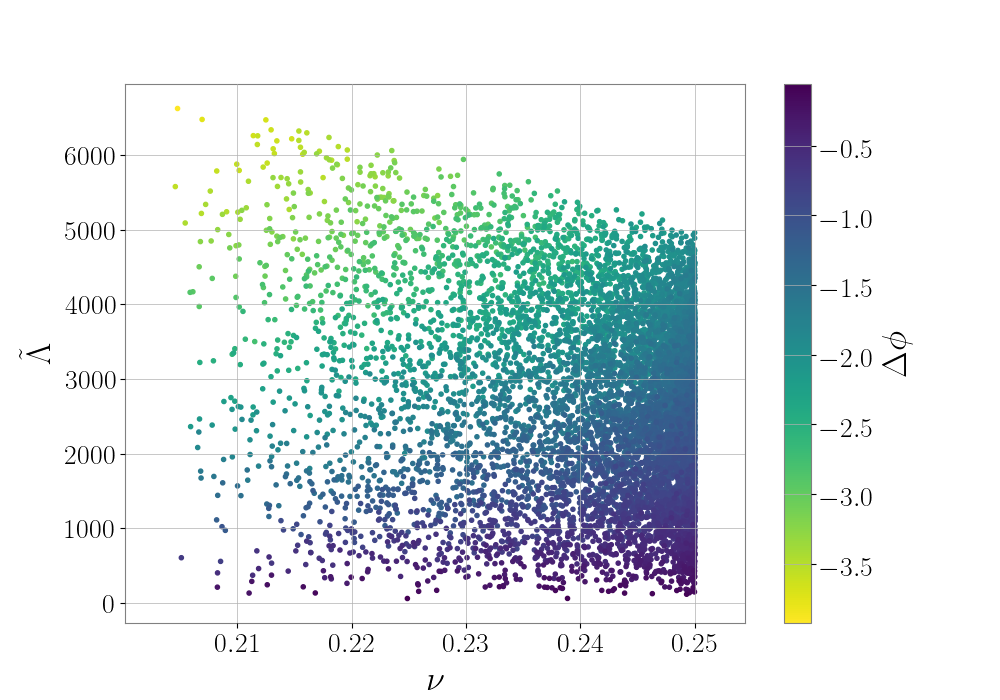}}
    \subfloat[]{\includegraphics[trim=0 0 50 50, clip, width=0.49\linewidth]{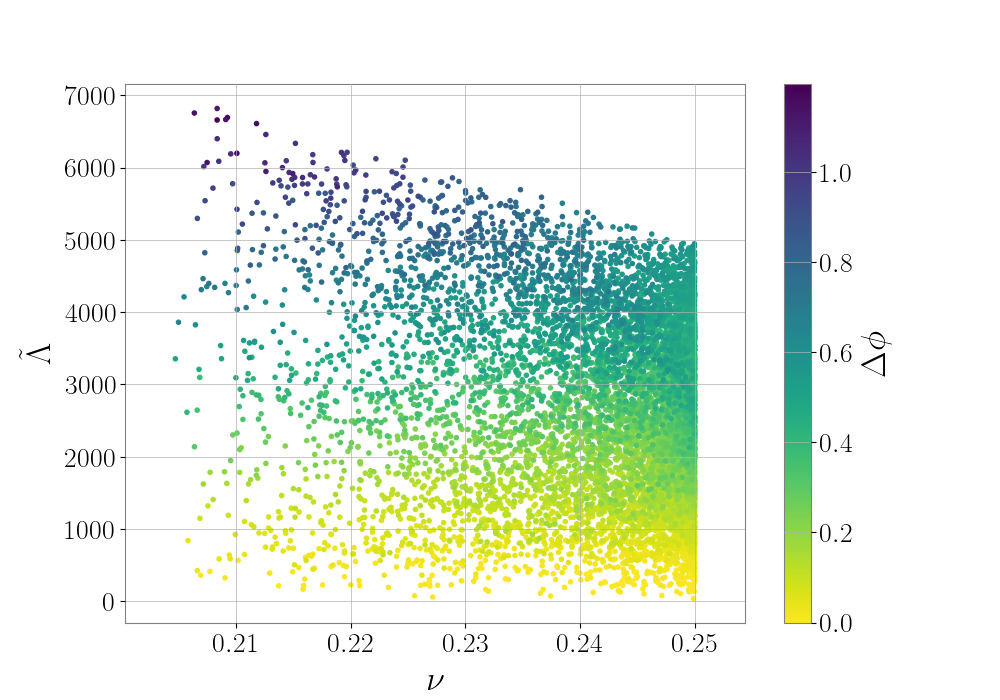}}
	\caption{Left: impact on the GW phase of the $\ell=3,4$ corrections in the binary interaction potential. The plot shows the phase contribution of these corrections for different binaries, identified by different values of the symmetric mass ratio and the reduced tidal parameter $(\nu,\tilde\Lambda)$, evolving from a GW frequency of $20\,{\rm Hz}$ to merger. Right: GW phase differences given by the fitting relations $\bar\Lambda_2(\bar\Lambda_{3,4})$ and those given by \cite{Yagi:2013awa}}
	\label{dphigw}
\end{figure*}

\begin{figure*}
	\centering
	\subfloat[]{\includegraphics[trim=30 0 30 30, clip, width=0.49\linewidth]{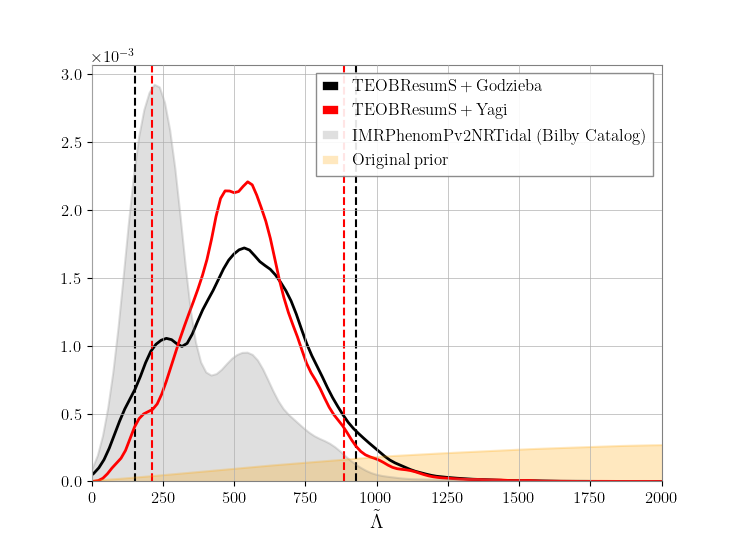}}
	\subfloat[]{\includegraphics[trim=10 0 10 30, clip, width=0.48\linewidth]{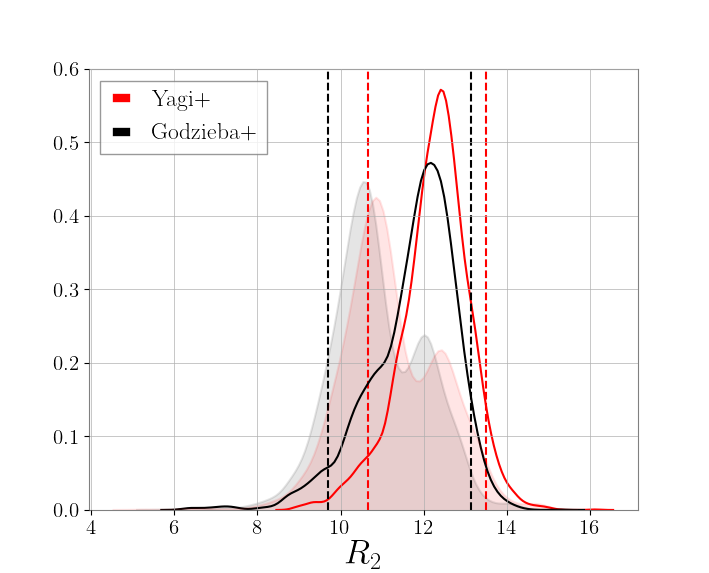}}
	\caption{Left panel: the marginalized one dimensional posterior distributions of $\tilde\Lambda$, recovered with the \TEOB{} approximant and the fits of Yagi (red) or those presented in this paper (black). We additionally compare our results with those obtained in \cite{Romero-Shaw:2020owr} (shaded gray area). Although statistical fluctuations are larger than any systematic effect due to the choice of quasi-universal relation, the use of the new phenomenological relations slightly increases the support at values of $\tilde\Lambda \lesssim 300$. Right panel: the radius of the lighter component of the binary, estimated through quasiuniversal relations from the reduced tidal parameter and mass distributions. 
	The shaded distributions correspond to the values obtained when mapping the {\tt bilby} catalog
	$\tilde\Lambda$ posteriors into radii values with either the new fits (gray) or the Yagi formulae. We find that the newly fitted coefficients for Eq. (\ref{C-lambda2 fit}) lead to lower radii values than those predicted by Yagi and Yunes. 
	}
	\label{RGW170817}
\end{figure*}

We now discuss the impact of the new quasiuniversal relations in GW modeling and PE using the state-of-art effective-one-body model \TEOB{}, that provides us with multipolar tidal waveforms for the full inspiral-merger regime~\cite{Bernuzzi:2015rla,Nagar:2018plt,Akcay:2018yyh}. Specifically, the tidal sector of \TEOB{} includes the $\ell=2,3,4$ gravitoelectric and the $\ell=2$ gravitomagnetic tidal terms in the binary interaction potential at the highest known post-Newtonian order, and additionally implements a resummation for the $\ell=2,3$ ($\ell=2$) gravitoelectric (magnetic) terms based on  gravitational-self-force results \cite{Bini:2012gu,Bini:2014zxa,Akcay:2018yyh}. The quasinuniversal relations among the tidal polarizability parameters with different $\ell$ are employed to obtain the octupolar and hexadecapolar from the quadrupolar parameters.

The importance of including the $\ell=3,4$ corrections in the waveform model is highlighted in the left panel of Fig.~\ref{dphigw}. The figure shows the contribution of the octupolar and hexadecapolar gravitoelectric terms for $10^3$ binaries that are identified by different values of the symmetric mass ratio and the reduced tidal parameter $\tilde \Lambda$. These tidal terms accelerate the inspiral and give an overall contribution of 0.5 radians to one cycle for a starting GW frequency of $20\,{\rm Hz}$, Cf. \cite{Lackey:2016krb}. The smallest phase differences are found for smaller reduced tidal parameters and equal masses binaries. The phase differences are accumulated from GW frequencies $\gtrsim 500\,{\rm Hz}$ (Cf. \cite{Damour:2012yf}), corresponding to the last orbits before merger. We remark that current differences between \TEOB{} and numerical-relativity waveforms are precisely of order 0.5-1 rad and are comparable to the numerical-relativity error~\cite{Bernuzzi:2015rla,Nagar:2018plt,Akcay:2018yyh}. Thus, this analysis indicates that an accurate modeling of higher multipoles in the binary interaction potential can be relevant to capture the merger waveform.

The accuracy of the fitting formulas for the $\Lambda_{3}$-$\Lambda_{2}$ and $\Lambda_{4}$-$\Lambda_{2}$ relations can impact significantly the GW phase. The GW phase differences induced by the use of the new fits of Eq.~\red{X} and those of \cite{Yagi:2013awa} is shown in the right panel of Fig.~\ref{dphigw}. Differences of order 10\% in the fitting formulas result in differences up to one radian in the GW (again we use an initial frequency of $20\,{\rm Hz}$). These differences are relevant for GW observations at signal-to-noise ratio ${\gtrsim}80$, at which the universal relations can, among other modeling choices, make a difference towards obtaining faithful waveform models \cite{Gamba:2020wgg}.



Finally, we perform Bayesian parameter estimation of GW170817 using \TEOB{} and the {\tt pbilby}~\cite{Smith:2019ucc, Romero-Shaw:2020owr} infrastructure with the {\tt dynesty}~\cite{2020MNRAS.493.3132S} sampler. Strain data is downloaded from the Gravitational Waves Open Science Center (GWOSC)~\cite{Abbott:2019ebz}, and the noise curves used are those provided in~\cite{LIGOScientific:2018mvr}. Our analysis is identical to the one performed in \cite{Gamba:2020wgg}, to which we refer for all the technical details, except for the the use of the quasiuniversal relations developed here. We just recall that we use a sampling rate of $2048\,{\rm Hz}$, that implies a cutoff maximum frequency of $1024\,{\rm Hz}$ for the analysis. This conservative choice ensures that no high-frequency systematic effect will affect our estimates \cite{Gamba:2020wgg}, and distinguishes our analysis from that of e.g~\cite{LIGOScientific:2018mvr, Romero-Shaw:2020owr}. In Fig. \ref{RGW170817} we report the marginalized one-dimensional posteriors for the reduced tidal parameter and the NS radius for the original run of \cite{Gamba:2020wgg} that used the Yagi relations (red inline) and for the one performed here with the new quasiuniversal relations (black inline). The two $\tilde\Lambda$ distributions are largely compatible, with the newly computed posteriors displaying slightly more support for low values ($\tilde\Lambda \lesssim 300$). This difference is negligible with respect to statistical fluctuations, but can nonetheless be understood by observing that the new fits predict stronger tidal effects than the ones of Yagi and Yunes. The reduced tidal parameter distributions can then, together with the mass ratio and component mass posteriors \footnote{Source and detector frame masses are linked by $m^{\rm source} = m^{\rm det}/(1+z)$, with $z=0.0099$ \cite{Abbott:2018wiz}}, be mapped into estimates of the radii of the stars. By combining Eq. (\ref{binary fit}) with the definition of the reduced tidal parameter and Eq. (\ref{C-lambda2 fit}) we estimate the distribution of the radius of the lighter star $R_2$. Using the coefficients collected in Tab. \ref{fitting parameters 2} and \ref{fitting parameters 3} we obtain $R_2^{\rm Godzieba+} = 11.9_{-2.1}^{+1.2}$ and $R_2^{\rm Yagi+} = 12.4_{-1.7}^{+1.2}$. While the two results lie well within the statistical uncertainty of the other, their difference ($ R_2^{\rm Godzieba, median} - R_2^{\rm Yagi, median} \sim 0.5$ km) can be fully attributed to quasiuniversal relations, in particular the $C$-$\Lambda_2$ fit. The discrepancy found, negligible for current events, will become very relevant when analyzing GW data from loud BNS events detected by advanced and third generation detectors. The source parameters recovered for such systems will be affected by small statistical fluctuations. A simple error propagation, applied to the fit of $R_{1.4 \Msun}$ of \cite{De:2018uhw}, gives \begin{equation} 
    \sigma_{R_{1.4\Msun}} = (11.2 \pm 0.2)\frac{\mathcal{M}}{4800}\Bigl(\frac{\tilde\Lambda}{800}\Bigr)^{-5/6}\sigma_{\tilde\Lambda}
\end{equation} where we assumed that the error on the chirp mass $\mathcal{M}$ is negligible. Therefore, $\sigma_{\tilde\Lambda} \sim 20$ (value expected for SNRs $\sim 300$) translates into $\sigma_{R_{1.4\Msun}}\sim 50\, {\rm m}$ for a $(\mathcal{M}, \tilde\Lambda = 1.18, 800)$ BNS system. This value amounts to approximately one tenth of the difference found above due to universal relations.

\section{\label{Conclusion}Conclusion}
We present updated fits to several universal relations between bulk properties of NSs relevant to current and future GW astronomy. The original fits can be found in \cite{Yagi:2013sva, Yagi:2015pkc, Yagi:2016qmr, Maselli:2013mva}. The updated fits are produced by sampling from a larger volume of the space of all possible NS EOSs not yet excluded by astronomical observation than had been done in the original fits. We sample the space of possible EOSs using an MCMC algorithm, which has a transition rate determined by three general physical constraints. Our results confirm and extends previously identified universal relations to a much larger set of EOSs.

First, we update fits to relations among three of the $l$-th order electric tidal deformabilities, $\Lambda_2$, $\Lambda_3$, and $\Lambda_4$. On the whole, the update decreases the relative errors of the fits by ${\sim}1\%$ compared to those of the original fits.

Next, we update the fit to the relation between the compactness $C$ and $\Lambda_2$. For small $C$ ($C < 0.25$), our fit has a relative error comparable to that of the original fit. However, for large $C$ ($C>0.25$), the error of our fit is as much as ${\sim}10\%$ smaller than that of the original fit. We can attribute this increase in accuracy at least in part to better sampling of EOSs that admit NSs with $C>0.3$. This is the largest improvement in accuracy among our updated fits.

Finally, we update the fit to the relation between the symmetric ($\Lambda_s$) and antisymmetric ($\Lambda_2$) combinations of $\Lambda_2$ for each star in a BNS. On the whole, the update decreases the relative error of the fit by ${\sim}1\%$ compared to that of the original fit. To demonstrate the improvements updating these fits make to GW analysis, we use the use the $C$-$\Lambda_2$ and $\Lambda_a$-$\Lambda_s$ relations to recover the radii of BNSs. We get ${\sim}0.5\%$ decreased relative error in the recovered radii using our updated fits versus using the original fits. This is due almost entirely to the drastic improvement in the accuracy of the $C$-$\Lambda_2$ fit.

We also discuss the implications of universal relations and the updated fits for GW waveform modelling and parameter estimation. Higher-order ($l>2$) multipole corrections in the waveform model are important for determining the GW phase of a merger. The $l=3,4$ corrections in particular contribute as much as 3.5 radians of accumulated dephasing at merger for a starting GW frequency of $20\,{\rm Hz}$. Thus, accurate modelling of $l>2$ multipoles is relevant for capturing the merger waveform faithfully. When using the original and updated multipole fits to recover $\Lambda_3$ and $\Lambda_4$ in waveform modelling, we find that the phase difference between the merger waveforms computed from each fit is of about 0.5 radians from $20\,{\rm Hz}$ to merger.

We perform a Bayesian parameter estimation of GW170817 using both the original and updated $C$-$\Lambda_2$ and $\Lambda_a$-$\Lambda_s$ fits, and we compare the distributions for the radius of the secondary in the merger yielded by each set of fits. The results from the updated fits are consistent with those of the original fits; however, the updated fits slightly favor a smaller radius, with the difference in the medians of the two distributions being ${\sim}0.5\,{\rm km}$. This is due almost entirely to the increased accuracy of the updated $C$-$\Lambda_2$ fit. Advanced and third generation GW detectors will be sensitive enough to measure NS radii to comparable accuracy. Thus, with increased sensitivity, the accuracy of fitting formulas for universal relations will become very relevant.

\begin{acknowledgments}
It is a pleasure to acknowledge B.~Sathyaprakash for discussions.
DR acknowledges support from the U.S. Department of Energy, Office of Science, Division of Nuclear Physics under Award Number(s) DE-SC0021177 and from the National Science Foundation under Grant No. PHY-2011725.
RG acknowledges support from the Deutsche Forschungsgemeinschaft (DFG) under Grant No. 406116891 within the Research Training Group RTG 2522/1.
SB acknowledges support by the EU H2020 under ERC Starting Grant, no.~BinGraSp-714626.  
Computations for this research were performed on the Pennsylvania State University's Institute for Computational and Data Sciences Advanced CyberInfrastructure (ICDS-ACI).
This research has made use of data, software and/or web tools obtained 
from the Gravitational Wave Open Science Center (\url{https://www.gw-openscience.org}), 
a service of LIGO Laboratory, the LIGO Scientific Collaboration and the 
Virgo Collaboration. LIGO is funded by the U.S. National Science Foundation. 
Virgo is funded by the French Centre National de Recherche Scientifique (CNRS), 
the Italian Istituto Nazionale della Fisica Nucleare (INFN) and the 
Dutch Nikhef, with contributions by Polish and Hungarian institutes.
\end{acknowledgments}



\bibliography{tidal_deform}

\begin{thebibliography}{52}%
\makeatletter
\providecommand \@ifxundefined [1]{%
 \@ifx{#1\undefined}
}%
\providecommand \@ifnum [1]{%
 \ifnum #1\expandafter \@firstoftwo
 \else \expandafter \@secondoftwo
 \fi
}%
\providecommand \@ifx [1]{%
 \ifx #1\expandafter \@firstoftwo
 \else \expandafter \@secondoftwo
 \fi
}%
\providecommand \natexlab [1]{#1}%
\providecommand \enquote  [1]{``#1''}%
\providecommand \bibnamefont  [1]{#1}%
\providecommand \bibfnamefont [1]{#1}%
\providecommand \citenamefont [1]{#1}%
\providecommand \href@noop [0]{\@secondoftwo}%
\providecommand \href [0]{\begingroup \@sanitize@url \@href}%
\providecommand \@href[1]{\@@startlink{#1}\@@href}%
\providecommand \@@href[1]{\endgroup#1\@@endlink}%
\providecommand \@sanitize@url [0]{\catcode `\\12\catcode `\$12\catcode
  `\&12\catcode `\#12\catcode `\^12\catcode `\_12\catcode `\%12\relax}%
\providecommand \@@startlink[1]{}%
\providecommand \@@endlink[0]{}%
\providecommand \url  [0]{\begingroup\@sanitize@url \@url }%
\providecommand \@url [1]{\endgroup\@href {#1}{\urlprefix }}%
\providecommand \urlprefix  [0]{URL }%
\providecommand \Eprint [0]{\href }%
\providecommand \doibase [0]{https://doi.org/}%
\providecommand \selectlanguage [0]{\@gobble}%
\providecommand \bibinfo  [0]{\@secondoftwo}%
\providecommand \bibfield  [0]{\@secondoftwo}%
\providecommand \translation [1]{[#1]}%
\providecommand \BibitemOpen [0]{}%
\providecommand \bibitemStop [0]{}%
\providecommand \bibitemNoStop [0]{.\EOS\space}%
\providecommand \EOS [0]{\spacefactor3000\relax}%
\providecommand \BibitemShut  [1]{\csname bibitem#1\endcsname}%
\let\auto@bib@innerbib\@empty
\bibitem [{\citenamefont {Özel}\ and\ \citenamefont
  {Freire}(2016)}]{Ozel:2016oaf}%
  \BibitemOpen
  \bibfield  {author} {\bibinfo {author} {\bibfnamefont {F.}~\bibnamefont
  {Özel}}\ and\ \bibinfo {author} {\bibfnamefont {P.}~\bibnamefont {Freire}},\
  }\bibfield  {title} {\bibinfo {title} {{Masses, Radii, and the Equation of
  State of Neutron Stars}},\ }\href
  {https://doi.org/10.1146/annurev-astro-081915-023322} {\bibfield  {journal}
  {\bibinfo  {journal} {Ann. Rev. Astron. Astrophys.}\ }\textbf {\bibinfo
  {volume} {54}},\ \bibinfo {pages} {401} (\bibinfo {year} {2016})},\ \Eprint
  {https://arxiv.org/abs/1603.02698} {arXiv:1603.02698 [astro-ph.HE]}
  \BibitemShut {NoStop}%
\bibitem [{\citenamefont {Lattimer}(2012)}]{Lattimer:2012nd}%
  \BibitemOpen
  \bibfield  {author} {\bibinfo {author} {\bibfnamefont {J.~M.}\ \bibnamefont
  {Lattimer}},\ }\bibfield  {title} {\bibinfo {title} {{The nuclear equation of
  state and neutron star masses}},\ }\href
  {https://doi.org/10.1146/annurev-nucl-102711-095018} {\bibfield  {journal}
  {\bibinfo  {journal} {Ann. Rev. Nucl. Part. Sci.}\ }\textbf {\bibinfo
  {volume} {62}},\ \bibinfo {pages} {485} (\bibinfo {year} {2012})},\ \Eprint
  {https://arxiv.org/abs/1305.3510} {arXiv:1305.3510 [nucl-th]} \BibitemShut
  {NoStop}%
\bibitem [{\citenamefont {Cromartie}\ \emph {et~al.}(2019)\citenamefont
  {Cromartie} \emph {et~al.}}]{Cromartie:2019kug}%
  \BibitemOpen
  \bibfield  {author} {\bibinfo {author} {\bibfnamefont {H.~T.}\ \bibnamefont
  {Cromartie}} \emph {et~al.},\ }\bibfield  {title} {\bibinfo {title}
  {{Relativistic Shapiro delay measurements of an extremely massive millisecond
  pulsar}},\ }\href {https://doi.org/10.1038/s41550-019-0880-2} {\bibfield
  {journal} {\bibinfo  {journal} {Nature Astron.}\ }\textbf {\bibinfo {volume}
  {4}},\ \bibinfo {pages} {72} (\bibinfo {year} {2019})},\ \Eprint
  {https://arxiv.org/abs/1904.06759} {arXiv:1904.06759 [astro-ph.HE]}
  \BibitemShut {NoStop}%
\bibitem [{\citenamefont {Steiner}\ \emph {et~al.}(2010)\citenamefont
  {Steiner}, \citenamefont {Lattimer},\ and\ \citenamefont
  {Brown}}]{Steiner:2010fz}%
  \BibitemOpen
  \bibfield  {author} {\bibinfo {author} {\bibfnamefont {A.~W.}\ \bibnamefont
  {Steiner}}, \bibinfo {author} {\bibfnamefont {J.~M.}\ \bibnamefont
  {Lattimer}},\ and\ \bibinfo {author} {\bibfnamefont {E.~F.}\ \bibnamefont
  {Brown}},\ }\bibfield  {title} {\bibinfo {title} {{The Equation of State from
  Observed Masses and Radii of Neutron Stars}},\ }\href
  {https://doi.org/10.1088/0004-637X/722/1/33} {\bibfield  {journal} {\bibinfo
  {journal} {Astrophys. J.}\ }\textbf {\bibinfo {volume} {722}},\ \bibinfo
  {pages} {33} (\bibinfo {year} {2010})},\ \Eprint
  {https://arxiv.org/abs/1005.0811} {arXiv:1005.0811 [astro-ph.HE]}
  \BibitemShut {NoStop}%
\bibitem [{\citenamefont {Steiner}\ \emph {et~al.}(2013)\citenamefont
  {Steiner}, \citenamefont {Lattimer},\ and\ \citenamefont
  {Brown}}]{Steiner:2012xt}%
  \BibitemOpen
  \bibfield  {author} {\bibinfo {author} {\bibfnamefont {A.~W.}\ \bibnamefont
  {Steiner}}, \bibinfo {author} {\bibfnamefont {J.~M.}\ \bibnamefont
  {Lattimer}},\ and\ \bibinfo {author} {\bibfnamefont {E.~F.}\ \bibnamefont
  {Brown}},\ }\bibfield  {title} {\bibinfo {title} {{The Neutron Star
  Mass-Radius Relation and the Equation of State of Dense Matter}},\ }\href
  {https://doi.org/10.1088/2041-8205/765/1/L5} {\bibfield  {journal} {\bibinfo
  {journal} {Astrophys. J. Lett.}\ }\textbf {\bibinfo {volume} {765}},\
  \bibinfo {pages} {L5} (\bibinfo {year} {2013})},\ \Eprint
  {https://arxiv.org/abs/1205.6871} {arXiv:1205.6871 [nucl-th]} \BibitemShut
  {NoStop}%
\bibitem [{\citenamefont {Ozel}\ \emph {et~al.}(2016)\citenamefont {Ozel},
  \citenamefont {Psaltis}, \citenamefont {Guver}, \citenamefont {Baym},
  \citenamefont {Heinke},\ and\ \citenamefont {Guillot}}]{Ozel:2015fia}%
  \BibitemOpen
  \bibfield  {author} {\bibinfo {author} {\bibfnamefont {F.}~\bibnamefont
  {Ozel}}, \bibinfo {author} {\bibfnamefont {D.}~\bibnamefont {Psaltis}},
  \bibinfo {author} {\bibfnamefont {T.}~\bibnamefont {Guver}}, \bibinfo
  {author} {\bibfnamefont {G.}~\bibnamefont {Baym}}, \bibinfo {author}
  {\bibfnamefont {C.}~\bibnamefont {Heinke}},\ and\ \bibinfo {author}
  {\bibfnamefont {S.}~\bibnamefont {Guillot}},\ }\bibfield  {title} {\bibinfo
  {title} {{The Dense Matter Equation of State from Neutron Star Radius and
  Mass Measurements}},\ }\href {https://doi.org/10.3847/0004-637X/820/1/28}
  {\bibfield  {journal} {\bibinfo  {journal} {Astrophys. J.}\ }\textbf
  {\bibinfo {volume} {820}},\ \bibinfo {pages} {28} (\bibinfo {year} {2016})},\
  \Eprint {https://arxiv.org/abs/1505.05155} {arXiv:1505.05155 [astro-ph.HE]}
  \BibitemShut {NoStop}%
\bibitem [{\citenamefont {Most}\ \emph {et~al.}(2018)\citenamefont {Most},
  \citenamefont {Weih}, \citenamefont {Rezzolla},\ and\ \citenamefont
  {Schaffner-Bielich}}]{Most:2018hfd}%
  \BibitemOpen
  \bibfield  {author} {\bibinfo {author} {\bibfnamefont {E.~R.}\ \bibnamefont
  {Most}}, \bibinfo {author} {\bibfnamefont {L.~R.}\ \bibnamefont {Weih}},
  \bibinfo {author} {\bibfnamefont {L.}~\bibnamefont {Rezzolla}},\ and\
  \bibinfo {author} {\bibfnamefont {J.}~\bibnamefont {Schaffner-Bielich}},\
  }\bibfield  {title} {\bibinfo {title} {{New constraints on radii and tidal
  deformabilities of neutron stars from GW170817}},\ }\href
  {https://doi.org/10.1103/PhysRevLett.120.261103} {\bibfield  {journal}
  {\bibinfo  {journal} {Phys. Rev. Lett.}\ }\textbf {\bibinfo {volume} {120}},\
  \bibinfo {pages} {261103} (\bibinfo {year} {2018})},\ \Eprint
  {https://arxiv.org/abs/1803.00549} {arXiv:1803.00549 [gr-qc]} \BibitemShut
  {NoStop}%
\bibitem [{\citenamefont {Capano}\ \emph {et~al.}(2020)\citenamefont {Capano},
  \citenamefont {Tews}, \citenamefont {Brown}, \citenamefont {Margalit},
  \citenamefont {De}, \citenamefont {Kumar}, \citenamefont {Brown},
  \citenamefont {Krishnan},\ and\ \citenamefont {Reddy}}]{Capano:2019eae}%
  \BibitemOpen
  \bibfield  {author} {\bibinfo {author} {\bibfnamefont {C.~D.}\ \bibnamefont
  {Capano}}, \bibinfo {author} {\bibfnamefont {I.}~\bibnamefont {Tews}},
  \bibinfo {author} {\bibfnamefont {S.~M.}\ \bibnamefont {Brown}}, \bibinfo
  {author} {\bibfnamefont {B.}~\bibnamefont {Margalit}}, \bibinfo {author}
  {\bibfnamefont {S.}~\bibnamefont {De}}, \bibinfo {author} {\bibfnamefont
  {S.}~\bibnamefont {Kumar}}, \bibinfo {author} {\bibfnamefont {D.~A.}\
  \bibnamefont {Brown}}, \bibinfo {author} {\bibfnamefont {B.}~\bibnamefont
  {Krishnan}},\ and\ \bibinfo {author} {\bibfnamefont {S.}~\bibnamefont
  {Reddy}},\ }\bibfield  {title} {\bibinfo {title} {{Stringent constraints on
  neutron-star radii from multimessenger observations and nuclear theory}},\
  }\href {https://doi.org/10.1038/s41550-020-1014-6} {\bibfield  {journal}
  {\bibinfo  {journal} {Nature Astron.}\ }\textbf {\bibinfo {volume} {4}},\
  \bibinfo {pages} {625} (\bibinfo {year} {2020})},\ \Eprint
  {https://arxiv.org/abs/1908.10352} {arXiv:1908.10352 [astro-ph.HE]}
  \BibitemShut {NoStop}%
\bibitem [{\citenamefont {Drischler}\ \emph {et~al.}(2020)\citenamefont
  {Drischler}, \citenamefont {Han}, \citenamefont {Lattimer}, \citenamefont
  {Prakash}, \citenamefont {Reddy},\ and\ \citenamefont
  {Zhao}}]{Drischler:2020fvz}%
  \BibitemOpen
  \bibfield  {author} {\bibinfo {author} {\bibfnamefont {C.}~\bibnamefont
  {Drischler}}, \bibinfo {author} {\bibfnamefont {S.}~\bibnamefont {Han}},
  \bibinfo {author} {\bibfnamefont {J.~M.}\ \bibnamefont {Lattimer}}, \bibinfo
  {author} {\bibfnamefont {M.}~\bibnamefont {Prakash}}, \bibinfo {author}
  {\bibfnamefont {S.}~\bibnamefont {Reddy}},\ and\ \bibinfo {author}
  {\bibfnamefont {T.}~\bibnamefont {Zhao}},\ }\bibfield  {title} {\bibinfo
  {title} {{Limiting masses and radii of neutron stars and their
  implications}},\ }\href@noop {} {\bibfield  {journal} {\bibinfo  {journal}
  {arXiv e-prints}\ } (\bibinfo {year} {2020})},\ \Eprint
  {https://arxiv.org/abs/2009.06441} {arXiv:2009.06441 [nucl-th]} \BibitemShut
  {NoStop}%
\bibitem [{\citenamefont {Greif}\ \emph {et~al.}(2019)\citenamefont {Greif},
  \citenamefont {Raaijmakers}, \citenamefont {Hebeler}, \citenamefont
  {Schwenk},\ and\ \citenamefont {Watts}}]{Greif:2018njt}%
  \BibitemOpen
  \bibfield  {author} {\bibinfo {author} {\bibfnamefont {S.}~\bibnamefont
  {Greif}}, \bibinfo {author} {\bibfnamefont {G.}~\bibnamefont {Raaijmakers}},
  \bibinfo {author} {\bibfnamefont {K.}~\bibnamefont {Hebeler}}, \bibinfo
  {author} {\bibfnamefont {A.}~\bibnamefont {Schwenk}},\ and\ \bibinfo {author}
  {\bibfnamefont {A.}~\bibnamefont {Watts}},\ }\bibfield  {title} {\bibinfo
  {title} {{Equation of state sensitivities when inferring neutron star and
  dense matter properties}},\ }\href {https://doi.org/10.1093/mnras/stz654}
  {\bibfield  {journal} {\bibinfo  {journal} {Mon. Not. Roy. Astron. Soc.}\
  }\textbf {\bibinfo {volume} {485}},\ \bibinfo {pages} {5363} (\bibinfo {year}
  {2019})},\ \Eprint {https://arxiv.org/abs/1812.08188} {arXiv:1812.08188
  [astro-ph.HE]} \BibitemShut {NoStop}%
\bibitem [{\citenamefont {Godzieba}\ \emph {et~al.}(2020)\citenamefont
  {Godzieba}, \citenamefont {Radice},\ and\ \citenamefont
  {Bernuzzi}}]{Godzieba:2020tjn}%
  \BibitemOpen
  \bibfield  {author} {\bibinfo {author} {\bibfnamefont {D.~A.}\ \bibnamefont
  {Godzieba}}, \bibinfo {author} {\bibfnamefont {D.}~\bibnamefont {Radice}},\
  and\ \bibinfo {author} {\bibfnamefont {S.}~\bibnamefont {Bernuzzi}},\
  }\bibfield  {title} {\bibinfo {title} {{On the maximum mass of neutron stars
  and GW190814}},\ }\href@noop {} {\bibfield  {journal} {\bibinfo  {journal}
  {arXiv e-prints}\ } (\bibinfo {year} {2020})},\ \Eprint
  {https://arxiv.org/abs/2007.10999} {arXiv:2007.10999 [astro-ph.HE]}
  \BibitemShut {NoStop}%
\bibitem [{\citenamefont {Annala}\ \emph {et~al.}(2018)\citenamefont {Annala},
  \citenamefont {Gorda}, \citenamefont {Kurkela},\ and\ \citenamefont
  {Vuorinen}}]{Annala:2017llu}%
  \BibitemOpen
  \bibfield  {author} {\bibinfo {author} {\bibfnamefont {E.}~\bibnamefont
  {Annala}}, \bibinfo {author} {\bibfnamefont {T.}~\bibnamefont {Gorda}},
  \bibinfo {author} {\bibfnamefont {A.}~\bibnamefont {Kurkela}},\ and\ \bibinfo
  {author} {\bibfnamefont {A.}~\bibnamefont {Vuorinen}},\ }\bibfield  {title}
  {\bibinfo {title} {{Gravitational-wave constraints on the neutron-star-matter
  Equation of State}},\ }\href {https://doi.org/10.1103/PhysRevLett.120.172703}
  {\bibfield  {journal} {\bibinfo  {journal} {Phys. Rev. Lett.}\ }\textbf
  {\bibinfo {volume} {120}},\ \bibinfo {pages} {172703} (\bibinfo {year}
  {2018})},\ \Eprint {https://arxiv.org/abs/1711.02644} {arXiv:1711.02644
  [astro-ph.HE]} \BibitemShut {NoStop}%
\bibitem [{\citenamefont {Annala}\ \emph {et~al.}(2020)\citenamefont {Annala},
  \citenamefont {Gorda}, \citenamefont {Kurkela}, \citenamefont {Nättilä},\
  and\ \citenamefont {Vuorinen}}]{Annala:2019puf}%
  \BibitemOpen
  \bibfield  {author} {\bibinfo {author} {\bibfnamefont {E.}~\bibnamefont
  {Annala}}, \bibinfo {author} {\bibfnamefont {T.}~\bibnamefont {Gorda}},
  \bibinfo {author} {\bibfnamefont {A.}~\bibnamefont {Kurkela}}, \bibinfo
  {author} {\bibfnamefont {J.}~\bibnamefont {Nättilä}},\ and\ \bibinfo
  {author} {\bibfnamefont {A.}~\bibnamefont {Vuorinen}},\ }\bibfield  {title}
  {\bibinfo {title} {{Evidence for quark-matter cores in massive neutron
  stars}},\ }\bibfield  {journal} {\bibinfo  {journal} {Nature Phys.}\ }\href
  {https://doi.org/10.1038/s41567-020-0914-9} {10.1038/s41567-020-0914-9}
  (\bibinfo {year} {2020}),\ \Eprint {https://arxiv.org/abs/1903.09121}
  {arXiv:1903.09121 [astro-ph.HE]} \BibitemShut {NoStop}%
\bibitem [{\citenamefont {De}\ \emph {et~al.}(2018)\citenamefont {De},
  \citenamefont {Finstad}, \citenamefont {Lattimer}, \citenamefont {Brown},
  \citenamefont {Berger},\ and\ \citenamefont {Biwer}}]{De:2018uhw}%
  \BibitemOpen
  \bibfield  {author} {\bibinfo {author} {\bibfnamefont {S.}~\bibnamefont
  {De}}, \bibinfo {author} {\bibfnamefont {D.}~\bibnamefont {Finstad}},
  \bibinfo {author} {\bibfnamefont {J.~M.}\ \bibnamefont {Lattimer}}, \bibinfo
  {author} {\bibfnamefont {D.~A.}\ \bibnamefont {Brown}}, \bibinfo {author}
  {\bibfnamefont {E.}~\bibnamefont {Berger}},\ and\ \bibinfo {author}
  {\bibfnamefont {C.~M.}\ \bibnamefont {Biwer}},\ }\bibfield  {title} {\bibinfo
  {title} {{Tidal Deformabilities and Radii of Neutron Stars from the
  Observation of GW170817}},\ }\href
  {https://doi.org/10.1103/PhysRevLett.121.091102} {\bibfield  {journal}
  {\bibinfo  {journal} {Phys. Rev. Lett.}\ }\textbf {\bibinfo {volume} {121}},\
  \bibinfo {pages} {091102} (\bibinfo {year} {2018})},\ \bibinfo {note}
  {[Erratum: Phys.Rev.Lett. 121, 259902 (2018)]},\ \Eprint
  {https://arxiv.org/abs/1804.08583} {arXiv:1804.08583 [astro-ph.HE]}
  \BibitemShut {NoStop}%
\bibitem [{\citenamefont {Hebeler}\ \emph {et~al.}(2013)\citenamefont
  {Hebeler}, \citenamefont {Lattimer}, \citenamefont {Pethick},\ and\
  \citenamefont {Schwenk}}]{Hebeler:2013nza}%
  \BibitemOpen
  \bibfield  {author} {\bibinfo {author} {\bibfnamefont {K.}~\bibnamefont
  {Hebeler}}, \bibinfo {author} {\bibfnamefont {J.}~\bibnamefont {Lattimer}},
  \bibinfo {author} {\bibfnamefont {C.}~\bibnamefont {Pethick}},\ and\ \bibinfo
  {author} {\bibfnamefont {A.}~\bibnamefont {Schwenk}},\ }\bibfield  {title}
  {\bibinfo {title} {{Equation of state and neutron star properties constrained
  by nuclear physics and observation}},\ }\href
  {https://doi.org/10.1088/0004-637X/773/1/11} {\bibfield  {journal} {\bibinfo
  {journal} {Astrophys. J.}\ }\textbf {\bibinfo {volume} {773}},\ \bibinfo
  {pages} {11} (\bibinfo {year} {2013})},\ \Eprint
  {https://arxiv.org/abs/1303.4662} {arXiv:1303.4662 [astro-ph.SR]}
  \BibitemShut {NoStop}%
\bibitem [{\citenamefont {Yagi}(2014)}]{Yagi:2013sva}%
  \BibitemOpen
  \bibfield  {author} {\bibinfo {author} {\bibfnamefont {K.}~\bibnamefont
  {Yagi}},\ }\bibfield  {title} {\bibinfo {title} {{Multipole Love
  Relations}},\ }\href {https://doi.org/10.1103/PhysRevD.89.043011} {\bibfield
  {journal} {\bibinfo  {journal} {Phys. Rev. D}\ }\textbf {\bibinfo {volume}
  {89}},\ \bibinfo {pages} {043011} (\bibinfo {year} {2014})},\ \bibinfo {note}
  {[Erratum: Phys. Rev. D 96, 129904 (2017), Erratum: Phys. Rev. D 97, 129901
  (2018)]},\ \Eprint {https://arxiv.org/abs/1311.0872} {arXiv:1311.0872
  [gr-qc]} \BibitemShut {NoStop}%
\bibitem [{\citenamefont {Yagi}\ and\ \citenamefont
  {Yunes}(2016)}]{Yagi:2015pkc}%
  \BibitemOpen
  \bibfield  {author} {\bibinfo {author} {\bibfnamefont {K.}~\bibnamefont
  {Yagi}}\ and\ \bibinfo {author} {\bibfnamefont {N.}~\bibnamefont {Yunes}},\
  }\bibfield  {title} {\bibinfo {title} {{Binary Love Relations}},\ }\href
  {https://doi.org/10.1088/0264-9381/33/13/13LT01} {\bibfield  {journal}
  {\bibinfo  {journal} {Class. Quant. Grav.}\ }\textbf {\bibinfo {volume}
  {33}},\ \bibinfo {pages} {13LT01} (\bibinfo {year} {2016})},\ \Eprint
  {https://arxiv.org/abs/1512.02639} {arXiv:1512.02639 [gr-qc]} \BibitemShut
  {NoStop}%
\bibitem [{\citenamefont {Yagi}\ and\ \citenamefont
  {Yunes}(2017{\natexlab{a}})}]{Yagi:2016qmr}%
  \BibitemOpen
  \bibfield  {author} {\bibinfo {author} {\bibfnamefont {K.}~\bibnamefont
  {Yagi}}\ and\ \bibinfo {author} {\bibfnamefont {N.}~\bibnamefont {Yunes}},\
  }\bibfield  {title} {\bibinfo {title} {{Approximate Universal Relations among
  Tidal Parameters for Neutron Star Binaries}},\ }\href
  {https://doi.org/10.1088/1361-6382/34/1/015006} {\bibfield  {journal}
  {\bibinfo  {journal} {Class. Quant. Grav.}\ }\textbf {\bibinfo {volume}
  {34}},\ \bibinfo {pages} {015006} (\bibinfo {year} {2017}{\natexlab{a}})},\
  \Eprint {https://arxiv.org/abs/1608.06187} {arXiv:1608.06187 [gr-qc]}
  \BibitemShut {NoStop}%
\bibitem [{\citenamefont {Abbott}\ \emph {et~al.}(2017)\citenamefont {Abbott}
  \emph {et~al.}}]{TheLIGOScientific:2017qsa}%
  \BibitemOpen
  \bibfield  {author} {\bibinfo {author} {\bibfnamefont {B.}~\bibnamefont
  {Abbott}} \emph {et~al.} (\bibinfo {collaboration} {LIGO Scientific,
  Virgo}),\ }\bibfield  {title} {\bibinfo {title} {{GW170817: Observation of
  Gravitational Waves from a Binary Neutron Star Inspiral}},\ }\href
  {https://doi.org/10.1103/PhysRevLett.119.161101} {\bibfield  {journal}
  {\bibinfo  {journal} {Phys. Rev. Lett.}\ }\textbf {\bibinfo {volume} {119}},\
  \bibinfo {pages} {161101} (\bibinfo {year} {2017})},\ \Eprint
  {https://arxiv.org/abs/1710.05832} {arXiv:1710.05832 [gr-qc]} \BibitemShut
  {NoStop}%
\bibitem [{\citenamefont {Abbott}\ \emph {et~al.}(2018)\citenamefont {Abbott}
  \emph {et~al.}}]{Abbott:2018exr}%
  \BibitemOpen
  \bibfield  {author} {\bibinfo {author} {\bibfnamefont {B.}~\bibnamefont
  {Abbott}} \emph {et~al.} (\bibinfo {collaboration} {LIGO Scientific,
  Virgo}),\ }\bibfield  {title} {\bibinfo {title} {{GW170817: Measurements of
  neutron star radii and equation of state}},\ }\href
  {https://doi.org/10.1103/PhysRevLett.121.161101} {\bibfield  {journal}
  {\bibinfo  {journal} {Phys. Rev. Lett.}\ }\textbf {\bibinfo {volume} {121}},\
  \bibinfo {pages} {161101} (\bibinfo {year} {2018})},\ \Eprint
  {https://arxiv.org/abs/1805.11581} {arXiv:1805.11581 [gr-qc]} \BibitemShut
  {NoStop}%
\bibitem [{\citenamefont {Yagi}\ and\ \citenamefont
  {Yunes}(2017{\natexlab{b}})}]{Yagi:2016bkt}%
  \BibitemOpen
  \bibfield  {author} {\bibinfo {author} {\bibfnamefont {K.}~\bibnamefont
  {Yagi}}\ and\ \bibinfo {author} {\bibfnamefont {N.}~\bibnamefont {Yunes}},\
  }\bibfield  {title} {\bibinfo {title} {{Approximate Universal Relations for
  Neutron Stars and Quark Stars}},\ }\href
  {https://doi.org/10.1016/j.physrep.2017.03.002} {\bibfield  {journal}
  {\bibinfo  {journal} {Phys. Rept.}\ }\textbf {\bibinfo {volume} {681}},\
  \bibinfo {pages} {1} (\bibinfo {year} {2017}{\natexlab{b}})},\ \Eprint
  {https://arxiv.org/abs/1608.02582} {arXiv:1608.02582 [gr-qc]} \BibitemShut
  {NoStop}%
\bibitem [{\citenamefont {Hinderer}\ \emph {et~al.}(2010)\citenamefont
  {Hinderer}, \citenamefont {Lackey}, \citenamefont {Lang},\ and\ \citenamefont
  {Read}}]{Hinderer:2009ca}%
  \BibitemOpen
  \bibfield  {author} {\bibinfo {author} {\bibfnamefont {T.}~\bibnamefont
  {Hinderer}}, \bibinfo {author} {\bibfnamefont {B.~D.}\ \bibnamefont
  {Lackey}}, \bibinfo {author} {\bibfnamefont {R.~N.}\ \bibnamefont {Lang}},\
  and\ \bibinfo {author} {\bibfnamefont {J.~S.}\ \bibnamefont {Read}},\
  }\bibfield  {title} {\bibinfo {title} {{Tidal deformability of neutron stars
  with realistic equations of state and their gravitational wave signatures in
  binary inspiral}},\ }\href {https://doi.org/10.1103/PhysRevD.81.123016}
  {\bibfield  {journal} {\bibinfo  {journal} {Phys. Rev. D}\ }\textbf {\bibinfo
  {volume} {81}},\ \bibinfo {pages} {123016} (\bibinfo {year} {2010})},\
  \Eprint {https://arxiv.org/abs/0911.3535} {arXiv:0911.3535 [astro-ph.HE]}
  \BibitemShut {NoStop}%
\bibitem [{\citenamefont {Damour}\ and\ \citenamefont
  {Nagar}(2009)}]{Damour:2009vw}%
  \BibitemOpen
  \bibfield  {author} {\bibinfo {author} {\bibfnamefont {T.}~\bibnamefont
  {Damour}}\ and\ \bibinfo {author} {\bibfnamefont {A.}~\bibnamefont {Nagar}},\
  }\bibfield  {title} {\bibinfo {title} {{Relativistic tidal properties of
  neutron stars}},\ }\href {https://doi.org/10.1103/PhysRevD.80.084035}
  {\bibfield  {journal} {\bibinfo  {journal} {Phys. Rev. D}\ }\textbf {\bibinfo
  {volume} {80}},\ \bibinfo {pages} {084035} (\bibinfo {year} {2009})},\
  \Eprint {https://arxiv.org/abs/0906.0096} {arXiv:0906.0096 [gr-qc]}
  \BibitemShut {NoStop}%
\bibitem [{\citenamefont {Postnikov}\ \emph {et~al.}(2010)\citenamefont
  {Postnikov}, \citenamefont {Prakash},\ and\ \citenamefont
  {Lattimer}}]{Postnikov:2010yn}%
  \BibitemOpen
  \bibfield  {author} {\bibinfo {author} {\bibfnamefont {S.}~\bibnamefont
  {Postnikov}}, \bibinfo {author} {\bibfnamefont {M.}~\bibnamefont {Prakash}},\
  and\ \bibinfo {author} {\bibfnamefont {J.~M.}\ \bibnamefont {Lattimer}},\
  }\bibfield  {title} {\bibinfo {title} {{Tidal Love Numbers of Neutron and
  Self-Bound Quark Stars}},\ }\href
  {https://doi.org/10.1103/PhysRevD.82.024016} {\bibfield  {journal} {\bibinfo
  {journal} {Phys. Rev. D}\ }\textbf {\bibinfo {volume} {82}},\ \bibinfo
  {pages} {024016} (\bibinfo {year} {2010})},\ \Eprint
  {https://arxiv.org/abs/1004.5098} {arXiv:1004.5098 [astro-ph.SR]}
  \BibitemShut {NoStop}%
\bibitem [{\citenamefont {Read}\ \emph {et~al.}(2009)\citenamefont {Read},
  \citenamefont {Lackey}, \citenamefont {Owen},\ and\ \citenamefont
  {Friedman}}]{Read:2008iy}%
  \BibitemOpen
  \bibfield  {author} {\bibinfo {author} {\bibfnamefont {J.~S.}\ \bibnamefont
  {Read}}, \bibinfo {author} {\bibfnamefont {B.~D.}\ \bibnamefont {Lackey}},
  \bibinfo {author} {\bibfnamefont {B.~J.}\ \bibnamefont {Owen}},\ and\
  \bibinfo {author} {\bibfnamefont {J.~L.}\ \bibnamefont {Friedman}},\
  }\bibfield  {title} {\bibinfo {title} {{Constraints on a phenomenologically
  parameterized neutron-star equation of state}},\ }\href
  {https://doi.org/10.1103/PhysRevD.79.124032} {\bibfield  {journal} {\bibinfo
  {journal} {Phys. Rev. D}\ }\textbf {\bibinfo {volume} {79}},\ \bibinfo
  {pages} {124032} (\bibinfo {year} {2009})},\ \Eprint
  {https://arxiv.org/abs/0812.2163} {arXiv:0812.2163 [astro-ph]} \BibitemShut
  {NoStop}%
\bibitem [{\citenamefont {Douchin}\ and\ \citenamefont
  {Haensel}(2001)}]{Douchin:2001sv}%
  \BibitemOpen
  \bibfield  {author} {\bibinfo {author} {\bibfnamefont {F.}~\bibnamefont
  {Douchin}}\ and\ \bibinfo {author} {\bibfnamefont {P.}~\bibnamefont
  {Haensel}},\ }\bibfield  {title} {\bibinfo {title} {{A unified equation of
  state of dense matter and neutron star structure}},\ }\href
  {https://doi.org/10.1051/0004-6361:20011402} {\bibfield  {journal} {\bibinfo
  {journal} {Astron. Astrophys.}\ }\textbf {\bibinfo {volume} {380}},\ \bibinfo
  {pages} {151} (\bibinfo {year} {2001})},\ \Eprint
  {https://arxiv.org/abs/astro-ph/0111092} {arXiv:astro-ph/0111092}
  \BibitemShut {NoStop}%
\bibitem [{\citenamefont {Rhoades}\ and\ \citenamefont
  {Ruffini}(1974)}]{Rhoades:1974fn}%
  \BibitemOpen
  \bibfield  {author} {\bibinfo {author} {\bibfnamefont {C.~E.}\ \bibnamefont
  {Rhoades}, \bibfnamefont {Jr.}}\ and\ \bibinfo {author} {\bibfnamefont
  {R.}~\bibnamefont {Ruffini}},\ }\bibfield  {title} {\bibinfo {title}
  {{Maximum mass of a neutron star}},\ }\href
  {https://doi.org/10.1103/PhysRevLett.32.324} {\bibfield  {journal} {\bibinfo
  {journal} {Phys. Rev. Lett.}\ }\textbf {\bibinfo {volume} {32}},\ \bibinfo
  {pages} {324} (\bibinfo {year} {1974})}\BibitemShut {NoStop}%
\bibitem [{\citenamefont {O'Boyle}\ \emph {et~al.}(2020)\citenamefont
  {O'Boyle}, \citenamefont {Markakis}, \citenamefont {Stergioulas},\ and\
  \citenamefont {Read}}]{OBoyle:2020qvf}%
  \BibitemOpen
  \bibfield  {author} {\bibinfo {author} {\bibfnamefont {M.~F.}\ \bibnamefont
  {O'Boyle}}, \bibinfo {author} {\bibfnamefont {C.}~\bibnamefont {Markakis}},
  \bibinfo {author} {\bibfnamefont {N.}~\bibnamefont {Stergioulas}},\ and\
  \bibinfo {author} {\bibfnamefont {J.~S.}\ \bibnamefont {Read}},\ }\bibfield
  {title} {\bibinfo {title} {{A Parametrized Equation of State for Neutron Star
  Matter with Continuous Sound Speed}},\ }\href@noop {} {\bibfield  {journal}
  {\bibinfo  {journal} {arXiv e-prints}\ } (\bibinfo {year} {2020})},\ \Eprint
  {https://arxiv.org/abs/2008.03342} {arXiv:2008.03342 [astro-ph.HE]}
  \BibitemShut {NoStop}%
\bibitem [{\citenamefont {Kanakis-Pegios}\ \emph {et~al.}(2020)\citenamefont
  {Kanakis-Pegios}, \citenamefont {Koliogiannis},\ and\ \citenamefont
  {Moustakidis}}]{Kanakis-Pegios:2020jnf}%
  \BibitemOpen
  \bibfield  {author} {\bibinfo {author} {\bibfnamefont {A.}~\bibnamefont
  {Kanakis-Pegios}}, \bibinfo {author} {\bibfnamefont {P.}~\bibnamefont
  {Koliogiannis}},\ and\ \bibinfo {author} {\bibfnamefont {C.}~\bibnamefont
  {Moustakidis}},\ }\bibfield  {title} {\bibinfo {title} {{Speed of sound
  constraints from tidal deformability of neutron stars}},\ }\href@noop {}
  {\bibfield  {journal} {\bibinfo  {journal} {arXiv e-prints}\ } (\bibinfo
  {year} {2020})},\ \Eprint {https://arxiv.org/abs/2007.13399}
  {arXiv:2007.13399 [nucl-th]} \BibitemShut {NoStop}%
\bibitem [{\citenamefont {Bernuzzi}\ and\ \citenamefont
  {Nagar}(2008)}]{Bernuzzi:2008fu}%
  \BibitemOpen
  \bibfield  {author} {\bibinfo {author} {\bibfnamefont {S.}~\bibnamefont
  {Bernuzzi}}\ and\ \bibinfo {author} {\bibfnamefont {A.}~\bibnamefont
  {Nagar}},\ }\bibfield  {title} {\bibinfo {title} {{Gravitational waves from
  pulsations of neutron stars described by realistic Equations of State}},\
  }\href {https://doi.org/10.1103/PhysRevD.78.024024} {\bibfield  {journal}
  {\bibinfo  {journal} {Phys. Rev. D}\ }\textbf {\bibinfo {volume} {78}},\
  \bibinfo {pages} {024024} (\bibinfo {year} {2008})},\ \Eprint
  {https://arxiv.org/abs/0803.3804} {arXiv:0803.3804 [gr-qc]} \BibitemShut
  {NoStop}%
\bibitem [{\citenamefont {Maselli}\ \emph {et~al.}(2013)\citenamefont
  {Maselli}, \citenamefont {Cardoso}, \citenamefont {Ferrari}, \citenamefont
  {Gualtieri},\ and\ \citenamefont {Pani}}]{Maselli:2013mva}%
  \BibitemOpen
  \bibfield  {author} {\bibinfo {author} {\bibfnamefont {A.}~\bibnamefont
  {Maselli}}, \bibinfo {author} {\bibfnamefont {V.}~\bibnamefont {Cardoso}},
  \bibinfo {author} {\bibfnamefont {V.}~\bibnamefont {Ferrari}}, \bibinfo
  {author} {\bibfnamefont {L.}~\bibnamefont {Gualtieri}},\ and\ \bibinfo
  {author} {\bibfnamefont {P.}~\bibnamefont {Pani}},\ }\bibfield  {title}
  {\bibinfo {title} {{Equation-of-state-independent relations in neutron
  stars}},\ }\href {https://doi.org/10.1103/PhysRevD.88.023007} {\bibfield
  {journal} {\bibinfo  {journal} {Phys. Rev. D}\ }\textbf {\bibinfo {volume}
  {88}},\ \bibinfo {pages} {023007} (\bibinfo {year} {2013})},\ \Eprint
  {https://arxiv.org/abs/1304.2052} {arXiv:1304.2052 [gr-qc]} \BibitemShut
  {NoStop}%
\bibitem [{\citenamefont {Kastaun}\ and\ \citenamefont
  {Ohme}(2019)}]{Kastaun:2019bxo}%
  \BibitemOpen
  \bibfield  {author} {\bibinfo {author} {\bibfnamefont {W.}~\bibnamefont
  {Kastaun}}\ and\ \bibinfo {author} {\bibfnamefont {F.}~\bibnamefont {Ohme}},\
  }\bibfield  {title} {\bibinfo {title} {{Finite tidal effects in GW170817:
  Observational evidence or model assumptions?}},\ }\href
  {https://doi.org/10.1103/PhysRevD.100.103023} {\bibfield  {journal} {\bibinfo
   {journal} {Phys. Rev. D}\ }\textbf {\bibinfo {volume} {100}},\ \bibinfo
  {pages} {103023} (\bibinfo {year} {2019})},\ \Eprint
  {https://arxiv.org/abs/1909.12718} {arXiv:1909.12718 [gr-qc]} \BibitemShut
  {NoStop}%
\bibitem [{\citenamefont {Chatziioannou}\ \emph {et~al.}(2018)\citenamefont
  {Chatziioannou}, \citenamefont {Haster},\ and\ \citenamefont
  {Zimmerman}}]{Chatziioannou:2018vzf}%
  \BibitemOpen
  \bibfield  {author} {\bibinfo {author} {\bibfnamefont {K.}~\bibnamefont
  {Chatziioannou}}, \bibinfo {author} {\bibfnamefont {C.-J.}\ \bibnamefont
  {Haster}},\ and\ \bibinfo {author} {\bibfnamefont {A.}~\bibnamefont
  {Zimmerman}},\ }\bibfield  {title} {\bibinfo {title} {{Measuring the neutron
  star tidal deformability with equation-of-state-independent relations and
  gravitational waves}},\ }\href {https://doi.org/10.1103/PhysRevD.97.104036}
  {\bibfield  {journal} {\bibinfo  {journal} {Phys. Rev. D}\ }\textbf {\bibinfo
  {volume} {97}},\ \bibinfo {pages} {104036} (\bibinfo {year} {2018})},\
  \Eprint {https://arxiv.org/abs/1804.03221} {arXiv:1804.03221 [gr-qc]}
  \BibitemShut {NoStop}%
\bibitem [{\citenamefont {Favata}(2014)}]{Favata:2013rwa}%
  \BibitemOpen
  \bibfield  {author} {\bibinfo {author} {\bibfnamefont {M.}~\bibnamefont
  {Favata}},\ }\bibfield  {title} {\bibinfo {title} {{Systematic parameter
  errors in inspiraling neutron star binaries}},\ }\href
  {https://doi.org/10.1103/PhysRevLett.112.101101} {\bibfield  {journal}
  {\bibinfo  {journal} {Phys. Rev. Lett.}\ }\textbf {\bibinfo {volume} {112}},\
  \bibinfo {pages} {101101} (\bibinfo {year} {2014})},\ \Eprint
  {https://arxiv.org/abs/1310.8288} {arXiv:1310.8288 [gr-qc]} \BibitemShut
  {NoStop}%
\bibitem [{\citenamefont {Flanagan}\ and\ \citenamefont
  {Hinderer}(2008)}]{Flanagan:2007ix}%
  \BibitemOpen
  \bibfield  {author} {\bibinfo {author} {\bibfnamefont {E.~E.}\ \bibnamefont
  {Flanagan}}\ and\ \bibinfo {author} {\bibfnamefont {T.}~\bibnamefont
  {Hinderer}},\ }\bibfield  {title} {\bibinfo {title} {{Constraining neutron
  star tidal Love numbers with gravitational wave detectors}},\ }\href
  {https://doi.org/10.1103/PhysRevD.77.021502} {\bibfield  {journal} {\bibinfo
  {journal} {Phys. Rev. D}\ }\textbf {\bibinfo {volume} {77}},\ \bibinfo
  {pages} {021502} (\bibinfo {year} {2008})},\ \Eprint
  {https://arxiv.org/abs/0709.1915} {arXiv:0709.1915 [astro-ph]} \BibitemShut
  {NoStop}%
\bibitem [{\citenamefont {Radice}\ \emph {et~al.}(2020)\citenamefont {Radice},
  \citenamefont {Bernuzzi},\ and\ \citenamefont {Perego}}]{Radice:2020ddv}%
  \BibitemOpen
  \bibfield  {author} {\bibinfo {author} {\bibfnamefont {D.}~\bibnamefont
  {Radice}}, \bibinfo {author} {\bibfnamefont {S.}~\bibnamefont {Bernuzzi}},\
  and\ \bibinfo {author} {\bibfnamefont {A.}~\bibnamefont {Perego}},\
  }\bibfield  {title} {\bibinfo {title} {{The Dynamics of Binary Neutron Star
  Mergers and of GW170817}},\ }\bibfield  {journal} {\bibinfo  {journal}
  {Annual Review of Nuclear and Particle Science}\ }\href
  {https://doi.org/10.1146/annurev-nucl-013120-114541}
  {10.1146/annurev-nucl-013120-114541} (\bibinfo {year} {2020}),\ \Eprint
  {https://arxiv.org/abs/2002.03863} {arXiv:2002.03863 [astro-ph.HE]}
  \BibitemShut {NoStop}%
\bibitem [{\citenamefont {Yagi}\ and\ \citenamefont
  {Yunes}(2013)}]{Yagi:2013awa}%
  \BibitemOpen
  \bibfield  {author} {\bibinfo {author} {\bibfnamefont {K.}~\bibnamefont
  {Yagi}}\ and\ \bibinfo {author} {\bibfnamefont {N.}~\bibnamefont {Yunes}},\
  }\bibfield  {title} {\bibinfo {title} {{I-Love-Q Relations in Neutron Stars
  and their Applications to Astrophysics, Gravitational Waves and Fundamental
  Physics}},\ }\href {https://doi.org/10.1103/PhysRevD.88.023009} {\bibfield
  {journal} {\bibinfo  {journal} {Phys. Rev. D}\ }\textbf {\bibinfo {volume}
  {88}},\ \bibinfo {pages} {023009} (\bibinfo {year} {2013})},\ \Eprint
  {https://arxiv.org/abs/1303.1528} {arXiv:1303.1528 [gr-qc]} \BibitemShut
  {NoStop}%
\bibitem [{\citenamefont {Romero-Shaw}\ \emph {et~al.}(2020)\citenamefont
  {Romero-Shaw} \emph {et~al.}}]{Romero-Shaw:2020owr}%
  \BibitemOpen
  \bibfield  {author} {\bibinfo {author} {\bibfnamefont {I.}~\bibnamefont
  {Romero-Shaw}} \emph {et~al.},\ }\bibfield  {title} {\bibinfo {title}
  {{Bayesian inference for compact binary coalescences with BILBY: Validation
  and application to the first LIGO--Virgo gravitational-wave transient
  catalogue}},\ }\bibfield  {journal} {\bibinfo  {journal} {arXiv e-prints}\
  }\href {https://doi.org/10.1093/mnras/staa2850} {10.1093/mnras/staa2850}
  (\bibinfo {year} {2020}),\ \Eprint {https://arxiv.org/abs/2006.00714}
  {arXiv:2006.00714 [astro-ph.IM]} \BibitemShut {NoStop}%
\bibitem [{\citenamefont {Bernuzzi}\ \emph {et~al.}(2015)\citenamefont
  {Bernuzzi}, \citenamefont {Dietrich},\ and\ \citenamefont
  {Nagar}}]{Bernuzzi:2015rla}%
  \BibitemOpen
  \bibfield  {author} {\bibinfo {author} {\bibfnamefont {S.}~\bibnamefont
  {Bernuzzi}}, \bibinfo {author} {\bibfnamefont {T.}~\bibnamefont {Dietrich}},\
  and\ \bibinfo {author} {\bibfnamefont {A.}~\bibnamefont {Nagar}},\ }\bibfield
   {title} {\bibinfo {title} {{Modeling the complete gravitational wave
  spectrum of neutron star mergers}},\ }\href
  {https://doi.org/10.1103/PhysRevLett.115.091101} {\bibfield  {journal}
  {\bibinfo  {journal} {Phys. Rev. Lett.}\ }\textbf {\bibinfo {volume} {115}},\
  \bibinfo {pages} {091101} (\bibinfo {year} {2015})},\ \Eprint
  {https://arxiv.org/abs/1504.01764} {arXiv:1504.01764 [gr-qc]} \BibitemShut
  {NoStop}%
\bibitem [{\citenamefont {Nagar}\ \emph {et~al.}(2019)\citenamefont {Nagar},
  \citenamefont {Messina}, \citenamefont {Rettegno}, \citenamefont {Bini},
  \citenamefont {Damour}, \citenamefont {Geralico}, \citenamefont {Akcay},\
  and\ \citenamefont {Bernuzzi}}]{Nagar:2018plt}%
  \BibitemOpen
  \bibfield  {author} {\bibinfo {author} {\bibfnamefont {A.}~\bibnamefont
  {Nagar}}, \bibinfo {author} {\bibfnamefont {F.}~\bibnamefont {Messina}},
  \bibinfo {author} {\bibfnamefont {P.}~\bibnamefont {Rettegno}}, \bibinfo
  {author} {\bibfnamefont {D.}~\bibnamefont {Bini}}, \bibinfo {author}
  {\bibfnamefont {T.}~\bibnamefont {Damour}}, \bibinfo {author} {\bibfnamefont
  {A.}~\bibnamefont {Geralico}}, \bibinfo {author} {\bibfnamefont
  {S.}~\bibnamefont {Akcay}},\ and\ \bibinfo {author} {\bibfnamefont
  {S.}~\bibnamefont {Bernuzzi}},\ }\bibfield  {title} {\bibinfo {title}
  {{Nonlinear-in-spin effects in effective-one-body waveform models of
  spin-aligned, inspiralling, neutron star binaries}},\ }\href
  {https://doi.org/10.1103/PhysRevD.99.044007} {\bibfield  {journal} {\bibinfo
  {journal} {Phys. Rev. D}\ }\textbf {\bibinfo {volume} {99}},\ \bibinfo
  {pages} {044007} (\bibinfo {year} {2019})},\ \Eprint
  {https://arxiv.org/abs/1812.07923} {arXiv:1812.07923 [gr-qc]} \BibitemShut
  {NoStop}%
\bibitem [{\citenamefont {Akcay}\ \emph {et~al.}(2019)\citenamefont {Akcay},
  \citenamefont {Bernuzzi}, \citenamefont {Messina}, \citenamefont {Nagar},
  \citenamefont {Ortiz},\ and\ \citenamefont {Rettegno}}]{Akcay:2018yyh}%
  \BibitemOpen
  \bibfield  {author} {\bibinfo {author} {\bibfnamefont {S.}~\bibnamefont
  {Akcay}}, \bibinfo {author} {\bibfnamefont {S.}~\bibnamefont {Bernuzzi}},
  \bibinfo {author} {\bibfnamefont {F.}~\bibnamefont {Messina}}, \bibinfo
  {author} {\bibfnamefont {A.}~\bibnamefont {Nagar}}, \bibinfo {author}
  {\bibfnamefont {N.}~\bibnamefont {Ortiz}},\ and\ \bibinfo {author}
  {\bibfnamefont {P.}~\bibnamefont {Rettegno}},\ }\bibfield  {title} {\bibinfo
  {title} {{Effective-one-body multipolar waveform for tidally interacting
  binary neutron stars up to merger}},\ }\href
  {https://doi.org/10.1103/PhysRevD.99.044051} {\bibfield  {journal} {\bibinfo
  {journal} {Phys. Rev. D}\ }\textbf {\bibinfo {volume} {99}},\ \bibinfo
  {pages} {044051} (\bibinfo {year} {2019})},\ \Eprint
  {https://arxiv.org/abs/1812.02744} {arXiv:1812.02744 [gr-qc]} \BibitemShut
  {NoStop}%
\bibitem [{\citenamefont {Bini}\ \emph {et~al.}(2012)\citenamefont {Bini},
  \citenamefont {Damour},\ and\ \citenamefont {Faye}}]{Bini:2012gu}%
  \BibitemOpen
  \bibfield  {author} {\bibinfo {author} {\bibfnamefont {D.}~\bibnamefont
  {Bini}}, \bibinfo {author} {\bibfnamefont {T.}~\bibnamefont {Damour}},\ and\
  \bibinfo {author} {\bibfnamefont {G.}~\bibnamefont {Faye}},\ }\bibfield
  {title} {\bibinfo {title} {{Effective action approach to higher-order
  relativistic tidal interactions in binary systems and their effective one
  body description}},\ }\href {https://doi.org/10.1103/PhysRevD.85.124034}
  {\bibfield  {journal} {\bibinfo  {journal} {Phys. Rev. D}\ }\textbf {\bibinfo
  {volume} {85}},\ \bibinfo {pages} {124034} (\bibinfo {year} {2012})},\
  \Eprint {https://arxiv.org/abs/1202.3565} {arXiv:1202.3565 [gr-qc]}
  \BibitemShut {NoStop}%
\bibitem [{\citenamefont {Bini}\ and\ \citenamefont
  {Damour}(2014)}]{Bini:2014zxa}%
  \BibitemOpen
  \bibfield  {author} {\bibinfo {author} {\bibfnamefont {D.}~\bibnamefont
  {Bini}}\ and\ \bibinfo {author} {\bibfnamefont {T.}~\bibnamefont {Damour}},\
  }\bibfield  {title} {\bibinfo {title} {{Gravitational self-force corrections
  to two-body tidal interactions and the effective one-body formalism}},\
  }\href {https://doi.org/10.1103/PhysRevD.90.124037} {\bibfield  {journal}
  {\bibinfo  {journal} {Phys. Rev. D}\ }\textbf {\bibinfo {volume} {90}},\
  \bibinfo {pages} {124037} (\bibinfo {year} {2014})},\ \Eprint
  {https://arxiv.org/abs/1409.6933} {arXiv:1409.6933 [gr-qc]} \BibitemShut
  {NoStop}%
\bibitem [{\citenamefont {Lackey}\ \emph {et~al.}(2017)\citenamefont {Lackey},
  \citenamefont {Bernuzzi}, \citenamefont {Galley}, \citenamefont {Meidam},\
  and\ \citenamefont {Van Den~Broeck}}]{Lackey:2016krb}%
  \BibitemOpen
  \bibfield  {author} {\bibinfo {author} {\bibfnamefont {B.~D.}\ \bibnamefont
  {Lackey}}, \bibinfo {author} {\bibfnamefont {S.}~\bibnamefont {Bernuzzi}},
  \bibinfo {author} {\bibfnamefont {C.~R.}\ \bibnamefont {Galley}}, \bibinfo
  {author} {\bibfnamefont {J.}~\bibnamefont {Meidam}},\ and\ \bibinfo {author}
  {\bibfnamefont {C.}~\bibnamefont {Van Den~Broeck}},\ }\bibfield  {title}
  {\bibinfo {title} {{Effective-one-body waveforms for binary neutron stars
  using surrogate models}},\ }\href
  {https://doi.org/10.1103/PhysRevD.95.104036} {\bibfield  {journal} {\bibinfo
  {journal} {Phys. Rev. D}\ }\textbf {\bibinfo {volume} {95}},\ \bibinfo
  {pages} {104036} (\bibinfo {year} {2017})},\ \Eprint
  {https://arxiv.org/abs/1610.04742} {arXiv:1610.04742 [gr-qc]} \BibitemShut
  {NoStop}%
\bibitem [{\citenamefont {Damour}\ \emph {et~al.}(2012)\citenamefont {Damour},
  \citenamefont {Nagar},\ and\ \citenamefont {Villain}}]{Damour:2012yf}%
  \BibitemOpen
  \bibfield  {author} {\bibinfo {author} {\bibfnamefont {T.}~\bibnamefont
  {Damour}}, \bibinfo {author} {\bibfnamefont {A.}~\bibnamefont {Nagar}},\ and\
  \bibinfo {author} {\bibfnamefont {L.}~\bibnamefont {Villain}},\ }\bibfield
  {title} {\bibinfo {title} {{Measurability of the tidal polarizability of
  neutron stars in late-inspiral gravitational-wave signals}},\ }\href
  {https://doi.org/10.1103/PhysRevD.85.123007} {\bibfield  {journal} {\bibinfo
  {journal} {Phys. Rev. D}\ }\textbf {\bibinfo {volume} {85}},\ \bibinfo
  {pages} {123007} (\bibinfo {year} {2012})},\ \Eprint
  {https://arxiv.org/abs/1203.4352} {arXiv:1203.4352 [gr-qc]} \BibitemShut
  {NoStop}%
\bibitem [{\citenamefont {Gamba}\ \emph {et~al.}(2020)\citenamefont {Gamba},
  \citenamefont {Breschi}, \citenamefont {Bernuzzi}, \citenamefont {Agathos},\
  and\ \citenamefont {Nagar}}]{Gamba:2020wgg}%
  \BibitemOpen
  \bibfield  {author} {\bibinfo {author} {\bibfnamefont {R.}~\bibnamefont
  {Gamba}}, \bibinfo {author} {\bibfnamefont {M.}~\bibnamefont {Breschi}},
  \bibinfo {author} {\bibfnamefont {S.}~\bibnamefont {Bernuzzi}}, \bibinfo
  {author} {\bibfnamefont {M.}~\bibnamefont {Agathos}},\ and\ \bibinfo {author}
  {\bibfnamefont {A.}~\bibnamefont {Nagar}},\ }\bibfield  {title} {\bibinfo
  {title} {{Waveform systematics in the gravitational-wave inference of tidal
  parameters and equation of state from binary neutron star signals}},\
  }\href@noop {} {\bibfield  {journal} {\bibinfo  {journal} {arXiv e-prints}\ }
  (\bibinfo {year} {2020})},\ \Eprint {https://arxiv.org/abs/2009.08467}
  {arXiv:2009.08467 [gr-qc]} \BibitemShut {NoStop}%
\bibitem [{\citenamefont {Smith}\ \emph {et~al.}(2020)\citenamefont {Smith},
  \citenamefont {Ashton}, \citenamefont {Vajpeyi},\ and\ \citenamefont
  {Talbot}}]{Smith:2019ucc}%
  \BibitemOpen
  \bibfield  {author} {\bibinfo {author} {\bibfnamefont {R.~J.}\ \bibnamefont
  {Smith}}, \bibinfo {author} {\bibfnamefont {G.}~\bibnamefont {Ashton}},
  \bibinfo {author} {\bibfnamefont {A.}~\bibnamefont {Vajpeyi}},\ and\ \bibinfo
  {author} {\bibfnamefont {C.}~\bibnamefont {Talbot}},\ }\bibfield  {title}
  {\bibinfo {title} {{Massively parallel Bayesian inference for transient
  gravitational-wave astronomy}},\ }\href
  {https://doi.org/10.1093/mnras/staa2483} {\bibfield  {journal} {\bibinfo
  {journal} {Mon. Not. Roy. Astron. Soc.}\ }\textbf {\bibinfo {volume} {498}},\
  \bibinfo {pages} {4492} (\bibinfo {year} {2020})},\ \Eprint
  {https://arxiv.org/abs/1909.11873} {arXiv:1909.11873 [gr-qc]} \BibitemShut
  {NoStop}%
\bibitem [{\citenamefont {Speagle}(2020)}]{2020MNRAS.493.3132S}%
  \BibitemOpen
  \bibfield  {author} {\bibinfo {author} {\bibfnamefont {J.~S.}\ \bibnamefont
  {Speagle}},\ }\bibfield  {title} {\bibinfo {title} {{dynesty: a dynamic
  nested sampling package for estimating Bayesian posteriors and evidences}},\
  }\href {https://doi.org/10.1093/mnras/staa278} {\bibfield  {journal}
  {\bibinfo  {journal} {Monthly Notices of the Royal Astronomical Society}\
  }\textbf {\bibinfo {volume} {493}},\ \bibinfo {pages} {3132} (\bibinfo {year}
  {2020})},\ \Eprint
  {https://arxiv.org/abs/https://academic.oup.com/mnras/article-pdf/493/3/3132/32890730/staa278.pdf}
  {https://academic.oup.com/mnras/article-pdf/493/3/3132/32890730/staa278.pdf}
  \BibitemShut {NoStop}%
\bibitem [{\citenamefont {Abbott}\ \emph
  {et~al.}(2019{\natexlab{a}})\citenamefont {Abbott} \emph
  {et~al.}}]{Abbott:2019ebz}%
  \BibitemOpen
  \bibfield  {author} {\bibinfo {author} {\bibfnamefont {R.}~\bibnamefont
  {Abbott}} \emph {et~al.} (\bibinfo {collaboration} {LIGO Scientific,
  Virgo}),\ }\bibfield  {title} {\bibinfo {title} {{Open data from the first
  and second observing runs of Advanced LIGO and Advanced Virgo}},\ }\href@noop
  {} {\bibfield  {journal} {\bibinfo  {journal} {arXiv e-prints}\ } (\bibinfo
  {year} {2019}{\natexlab{a}})},\ \Eprint {https://arxiv.org/abs/1912.11716}
  {arXiv:1912.11716 [gr-qc]} \BibitemShut {NoStop}%
\bibitem [{\citenamefont {Abbott}\ \emph
  {et~al.}(2019{\natexlab{b}})\citenamefont {Abbott} \emph
  {et~al.}}]{LIGOScientific:2018mvr}%
  \BibitemOpen
  \bibfield  {author} {\bibinfo {author} {\bibfnamefont {B.}~\bibnamefont
  {Abbott}} \emph {et~al.} (\bibinfo {collaboration} {LIGO Scientific,
  Virgo}),\ }\bibfield  {title} {\bibinfo {title} {{GWTC-1: A
  Gravitational-Wave Transient Catalog of Compact Binary Mergers Observed by
  LIGO and Virgo during the First and Second Observing Runs}},\ }\href
  {https://doi.org/10.1103/PhysRevX.9.031040} {\bibfield  {journal} {\bibinfo
  {journal} {Phys. Rev. X}\ }\textbf {\bibinfo {volume} {9}},\ \bibinfo {pages}
  {031040} (\bibinfo {year} {2019}{\natexlab{b}})},\ \Eprint
  {https://arxiv.org/abs/1811.12907} {arXiv:1811.12907 [astro-ph.HE]}
  \BibitemShut {NoStop}%
\bibitem [{Note1()}]{Note1}%
  \BibitemOpen
  \bibinfo {note} {Source and detector frame masses are linked by $m^{\protect
  \rm source} = m^{\protect \rm det}/(1+z)$, with $z=0.0099$ \cite
  {Abbott:2018wiz}}\BibitemShut {NoStop}%
\bibitem [{\citenamefont {Abbott}\ \emph
  {et~al.}(2019{\natexlab{c}})\citenamefont {Abbott} \emph
  {et~al.}}]{Abbott:2018wiz}%
  \BibitemOpen
  \bibfield  {author} {\bibinfo {author} {\bibfnamefont {B.}~\bibnamefont
  {Abbott}} \emph {et~al.} (\bibinfo {collaboration} {LIGO Scientific,
  Virgo}),\ }\bibfield  {title} {\bibinfo {title} {{Properties of the binary
  neutron star merger GW170817}},\ }\href
  {https://doi.org/10.1103/PhysRevX.9.011001} {\bibfield  {journal} {\bibinfo
  {journal} {Phys. Rev. X}\ }\textbf {\bibinfo {volume} {9}},\ \bibinfo {pages}
  {011001} (\bibinfo {year} {2019}{\natexlab{c}})},\ \Eprint
  {https://arxiv.org/abs/1805.11579} {arXiv:1805.11579 [gr-qc]} \BibitemShut
  {NoStop}%
\end{thebibliography}%


%

\end{document}